\documentclass[11pt]{article}

\usepackage[final]{acl}

\usepackage{times}
\usepackage{latexsym}
\usepackage[T1]{fontenc}
\usepackage{url}
\usepackage[utf8]{inputenc}

\usepackage{microtype}
\usepackage{algpseudocode}
\usepackage{inconsolata}
\usepackage{threeparttable}
\usepackage{graphicx}
\usepackage{hyperref}
\usepackage{url}

\usepackage{graphicx}
\usepackage[table]{xcolor}
\usepackage{amsmath}
\usepackage{amssymb}
\usepackage{booktabs}
\usepackage{pifont}
\usepackage{float}
\usepackage{subcaption}
\usepackage{booktabs}
\usepackage{multirow}
\usepackage{algorithm} 
\usepackage{xspace}
\usepackage{hyperref}
\usepackage{caption}
\usepackage{array}
\usepackage{placeins}
\usepackage{wrapfig}
\newcolumntype{L}[1]{>{\centering\arraybackslash}m{#1}}
\newcommand{\modelname}{Ours\xspace}

\usepackage{colortbl,xcolor}
\definecolor{blockyellow}{RGB}{255, 250, 220}
\definecolor{blockblue}{RGB}{240, 245, 255}
\definecolor{blockgreen}{RGB}{235, 250, 235}
\definecolor{lightblue}{RGB}{173, 216, 230}
\definecolor{lightgreen}{RGB}{144, 238, 144}
\definecolor{lightpurple}{RGB}{216, 191, 216}
\definecolor{lightred}{RGB}{255, 182, 193}

\definecolor{lightorange}{RGB}{191, 224, 255}
\definecolor{orange}{RGB}{120, 180, 240}
\title{AffectCodec: Emotion-Preserving Neural Speech Codec for Expressive Speech Modeling}

\usepackage{amssymb}

\author{
Jiacheng Shi$^{\spadesuit}$, Hongfei Du$^{\spadesuit}$, Xinyuan Song$^{\clubsuit}$, Y.~Alicia Hong$^{\heartsuit}$\\
\textbf{Yanfu Zhang$^{\spadesuit}$, Ye Gao$^{\spadesuit}$} \\
$^{\spadesuit}$College of William \& Mary,
$^{\clubsuit}$Emory University,
$^{\heartsuit}$George Mason University \\
\texttt{\small \{jshi12, hdu02, yzhang105, ygao18\}@wm.edu, xinyuan.song@emory.edu, yhong22@gmu.edu}
}

\begin{document}
\maketitle
\begin{abstract}

Neural speech codecs provide discrete representations for speech language models, but emotional cues are often degraded during quantization. Existing codecs mainly optimize acoustic reconstruction, leaving emotion expressiveness insufficiently modeled at the representation level. We propose an emotion-guided neural speech codec that explicitly preserves emotional information while maintaining semantic fidelity and prosodic naturalness. Our framework combines emotion–semantic guided latent modulation, relation-preserving emotional–semantic distillation, and emotion-weighted semantic alignment to retain emotionally salient cues under compression. Extensive evaluations across speech reconstruction, emotion recognition, and downstream text to speech generation demonstrate improved emotion consistency and perceptual quality without sacrificing content accuracy.
\end{abstract}

\section{Introduction}

Recent advances in large language models have rapidly extended to speech generation, enabling zero-shot text-to-speech~\cite{zero_tts_wang2023neural}, conversational voice agents~\cite{voice_agent_zeng2024glm}, and cross-modal audio reasoning~\cite{shi2025plug}. A key enabler of this progress is the neural speech codec~\cite{encodec_defossez2022high, zeghidour2021soundstream}, which converts continuous waveforms into discrete representations. By transforming high-rate acoustic signals into compact symbol sequences at fixed frame rates, neural codecs bridge the gap between continuous speech and token-based sequence learners, making large-scale training and inference computationally tractable.

As neural codecs are increasingly adopted as the discrete representation layer for speech language models, their ability to preserve emotional information has emerged as a critical concern. Recent benchmark studies~\cite{codec_superb_wu2024codec} show that, despite strong performance in reconstructing linguistic content and speaker characteristics, modern neural codecs exhibit substantial variation in emotion preservation. Large-scale evaluations consistently report degradation in downstream emotion recognition when speech is resynthesized through codecs, with sensitivity to bitrate, architecture, and training data. More fine-grained analyses~\cite{emocodec_ren2024emo} further indicate that subtle and expressive emotional cues are particularly vulnerable to distortion, even in state-of-the-art neural codecs, leading to a measurable loss of emotion integrity and expressiveness relative to original speech. Notably, such degradation often occurs despite high overall reconstruction quality, suggesting that emotional information is more fragile than other speech attributes. Taken together, these findings imply that existing neural codecs largely preserve emotion as an implicit byproduct of compression, rather than explicitly modeling emotional expressiveness or its interaction with prosody and semantic content during representation learning. These observations naturally motivate the following research question:
\emph{How can a neural speech codec preserve emotion integrity and expressiveness while simultaneously maintaining prosodic naturalness and semantic fidelity under discrete representation?}

To address this question, we propose an emotion-guided neural speech codec that reconsiders the optimization priorities of discrete speech representation learning. Unlike prior neural codecs that primarily emphasize acoustic reconstruction~\cite{encodec_defossez2022high, zeghidour2021soundstream} or semantic preservation~\cite{x-codec_ye2025codec, ye2025llasa} and implicitly retain emotional information, our approach elevates emotion preservation to a primary modeling objective, while jointly maintaining prosodic naturalness and semantic fidelity. Our framework consists of three complementary stages.
\textbf{(i) Emotion--Semantic Guided Latent Modulation} conditions acoustic latent representations on emotion- and semantics-related cues, enriching encoded features with affectively salient information prior to quantization.
\textbf{(ii) Relation-Preserving Emotional--Semantic Distillation} encourages the codec to retain emotion-related relational structure during representation transformation, mitigating affective degradation introduced by discrete quantization.
\textbf{(iii) Emotion-Weighted Semantic Alignment} further reinforces the association between discrete tokens and emotionally expressive content by adaptively weighting semantic alignment according to emotional salience.
Together, these stages form a unified codec framework that explicitly prioritizes emotional expressiveness while preserving the structural constraints required for fluent prosody and accurate semantic content. Our contributions can be summarized as follows:
\begin{itemize}
    \item \textbf{Conceptual Contribution.} We reconceptualize emotion preservation in neural speech codecs from a downstream evaluation concern to a core representation learning objective, explicitly treating emotional expressiveness as a primary optimization target rather than a post-hoc byproduct of acoustic reconstruction.
    \item \textbf{Methodological Contribution.} To the best of our knowledge, we present the first emotion-guided neural speech codec that addresses emotion degradation through a unified three-stage framework, integrating emotion--semantic guided latent, relation-preserving emotional--semantic distillation, and emotion-weighted semantic alignment.
    \item \textbf{Experimental Contribution.} Extensive experiments demonstrate that our approach improves reconstruction quality and representation effectiveness, and achieves strong emotion-related performance on the EMO-SUPERB and Codec-SUPERB benchmarks, as well as in zero-shot speech synthesis.
\end{itemize}
\begin{figure*}[htbp]
    \centering
    %\vspace{-14mm}
    \vspace{-8mm}
    % 调整宽度比例，例如 0.8\textwidth 表示占页面宽度的 80%
    \includegraphics[width=1\textwidth]{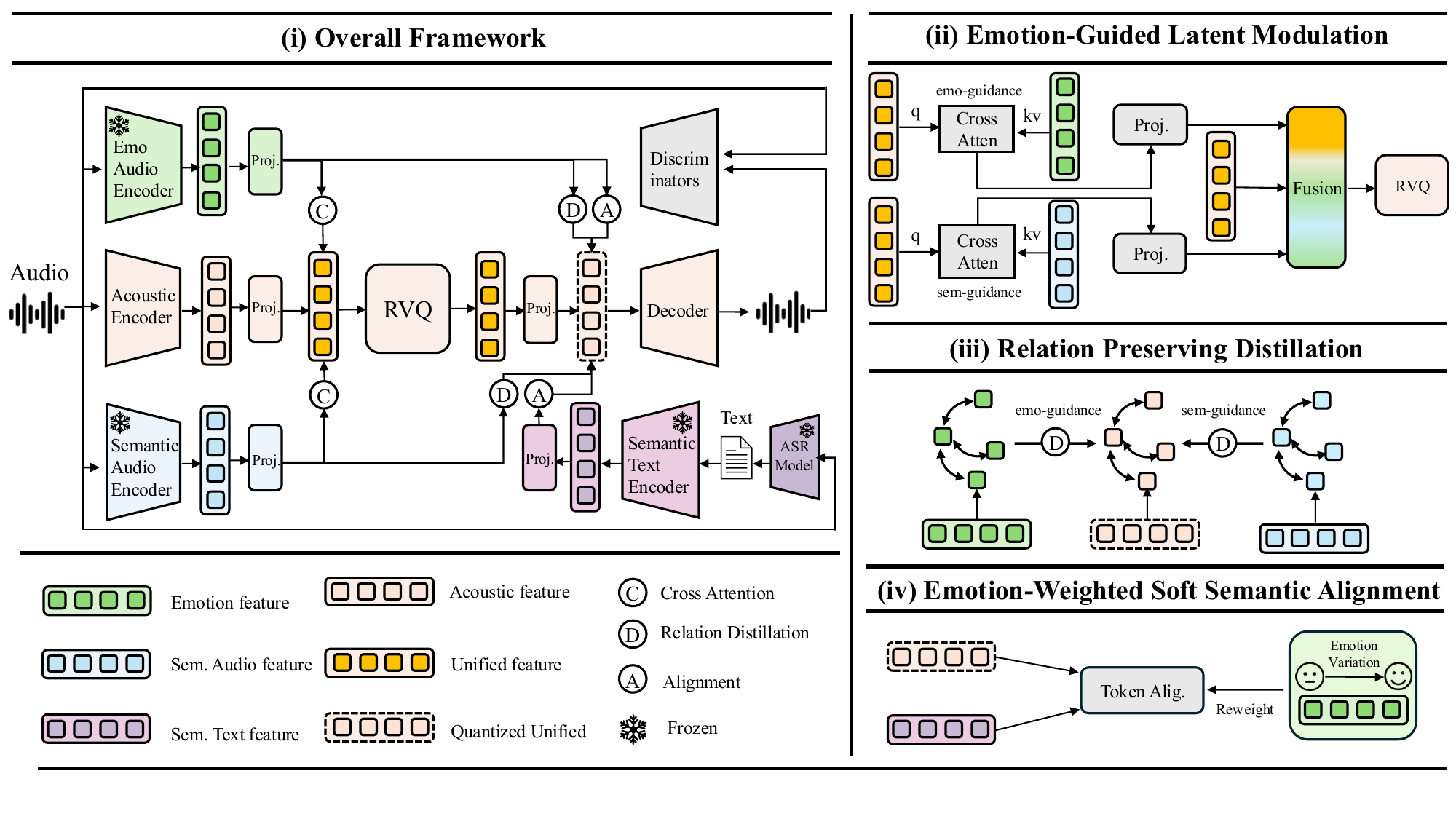}
    \vspace{-10mm}
    \caption{Overview of the proposed emotion-guided neural speech codec.
The codec encodes input speech into discrete acoustic representations via
residual vector quantization (RVQ) and incorporates emotion- and
semantic-aware mechanisms to preserve emotional expressiveness. Specifically,
it integrates (ii) emotion-guided latent modulation, which injects affective and
semantic cues into acoustic latents prior to quantization, (iii)
relation-preserving distillation, which constrains discrete representations to
retain relational structure from emotion and semantic spaces, and (iv)
emotion-weighted semantic alignment, which aligns quantized tokens with
textual semantics while emphasizing emotionally salient regions to maintain
semantic fidelity and prosodic naturalness.}
    \label{fig:main}
    \vspace{-2mm}
\end{figure*}
\section{Related Work}
We investigate prior work on Neural Speech Codecs and Discrete Audio Representation (Appendix~\ref{apx:related_neural_codec}) and Emotion-Aware Speech Representation Learning (Appendix~\ref{apx:emotion_related_work}), as these two research directions form the foundation for modeling discrete speech representations and preserving affective information in generative pipelines.
\section{Methods}
\subsection{Representation Backbone and Guidance}
As shown in Figure~\ref{fig:main}, discrete latent units provide a compact and temporally aligned representation for codec-based speech modeling following~\cite{encodec_defossez2022high,xin2024speechtokenizer}. 
Given an input waveform $\mathbf{x}$, we apply an acoustic encoder $E_{\mathrm{a}}$ to obtain a continuous latent sequence $\mathbf{A} = \{\mathbf{a}_t\}_{t=1}^{T'}$, where $\mathbf{a}_t \in \mathbb{R}^{D}$ denotes the frame-level representation and $T'$ is the number of encoded frames. These latents capture fine-grained spectral and prosodic information and form the basis for subsequent emotion-guided refinement. To obtain discrete representations, we adopt a residual vector quantization (RVQ) module with $K$ sequential codebook layers. At each layer $k$, the quantizer selects a codebook index sequence $\{q^{(k)}_t\}_{t=1}^{T'}$ and maps it to the corresponding embedding sequence $\mathbf{Q}^{(k)} \in \mathbb{R}^{T' \times D}$. 
The residual structure allows successive layers to model remaining quantization errors, yielding a refined approximation of the encoder latents. After $K$ layers, we obtain the full discrete acoustic representation $\mathbf{Q}^{(1:K)}$. Table~\ref{tab:motivation-efficiency} reports a comparison with prior codecs.

\noindent\textbf{Speech Emotion representation.}
We extract emotion-related features using a frozen emotion recognition model $G_{\mathrm{emo}}$, which maps the input waveform $\mathbf{x}$ to a sequence of frame-level embeddings $\mathbf{E} = \{\mathbf{e}_t\}_{t=1}^{T'}$, where $\mathbf{e}_t \in \mathbb{R}^{D_{\mathrm{e}}}$. 
These embeddings provide emotion guidance for subsequent representation learning. When multiple hidden layers are available, we aggregate the layer-wise representations by averaging,
$\mathbf{e}_t = \frac{1}{L_{\mathrm{e}}} \sum_{l=1}^{L_{\mathrm{e}}} \mathbf{h}^{(l)}_{\mathrm{emo}, t}$,
yielding a stable emotion embedding sequence $\mathbf{E}$ that serves as an emotion prior.

\noindent\textbf{Audio Semantic representation.}
We extract semantic audio features using a frozen self-supervised speech model $H$. 
Given the input waveform $\mathbf{x}$, the model produces a sequence of frame-level hidden states, which we aggregate across layers to obtain a sequence of semantic audio embeddings $\mathbf{S} = \{\mathbf{s}_t\}_{t=1}^{T'}$, where $\mathbf{s}_t \in \mathbb{R}^{D_{\mathrm{s}}}$. Specifically, when multiple hidden layers are available, we compute each semantic embedding by layer averaging,
$\mathbf{s}_t = \frac{1}{L} \sum_{\ell=1}^{L} \mathbf{h}^{(\ell)}_{t}$,
yielding a stable semantic representation sequence $\mathbf{S}$ that provides high-level semantic guidance for subsequent modules.

\noindent\textbf{Textual semantic representation.}
We extract textual semantic features using a frozen automatic speech recognition (ASR) model followed by a pre-trained language encoder. 
Given the input waveform $\mathbf{x}$, the ASR model produces a token sequence, which is then processed by the language encoder to obtain a sequence of token-level embeddings $\mathbf{C} = \{\mathbf{c}_t\}_{t=1}^{T}$, where $\mathbf{c}_t \in \mathbb{R}^{D_{\mathrm{c}}}$. When multiple transformer layers are available, we aggregate the layer-wise representations by averaging,
$\mathbf{c}_t = \frac{1}{L} \sum_{\ell=1}^{L} \mathbf{h}^{(\ell)}_{t}$,
yielding a stable textual semantic representation sequence $\mathbf{C}$ that provides linguistic guidance for subsequent modules.

\subsection{Emotion-Guided Optimization}
Standard neural speech codecs primarily optimize acoustic reconstruction~\cite{encodec_defossez2022high,xin2024speechtokenizer} or semantic preservation~\cite{x-codec_ye2025codec,ye2025llasa}, leaving emotional information implicitly encoded and vulnerable to degradation during quantization. This indicates that existing representations lack explicitly mechanisms to model emotional expressiveness or its interaction with acoustic and semantic structure. To address this limitation, we propose an emotion-guided optimization framework with three stages: \emph{Emotion-Semantic Guided Latent} (\S\ref{stage:1}), which injects emotion- and semantic-aware signals into acoustic latents prior to quantization; \emph{Relation-Preserving Emotional–Semantic Distillation} (\S\ref{stage:2}), which preserves emotion-related relational structure; and \emph{Emotion-Weighted Semantic Alignment} (\S\ref{stage:3}), which strengthens the association between discrete
tokens and emotionally expressive semantics.

\subsubsection{Emotion--Semantic Guided Latent}
\label{stage:1}
Given frame-level acoustic latents $\mathbf{Z}=\{\mathbf{z}_t\}_{t=1}^{T'}$, we incorporate emotion and semantic guidance prior to quantization. 
Specifically, we extract emotion embeddings $\mathbf{E}=\{\mathbf{e}_t\}_{t=1}^{T'}$ using a frozen emotion encoder and semantic audio embeddings $\mathbf{S}=\{\mathbf{s}_t\}_{t=1}^{T'}$ using a pretrained self-supervised speech model. 
All representations are projected into a shared interaction space via linear transformations, yielding $\tilde{\mathbf{z}}_t = W_a \mathbf{z}_t$,
$\tilde{\mathbf{e}}_t = W_e \mathbf{e}_t$,
and $\tilde{\mathbf{s}}_t = W_s \mathbf{s}_t$,
where $W_a \in \mathbb{R}^{D \times D}$,
$W_e \in \mathbb{R}^{D \times D_{\mathrm{e}}}$,
and $W_s \in \mathbb{R}^{D \times D_{\mathrm{s}}}$. We use acoustic features as queries in cross-attention over projected emotion and semantic sequences, where $\tilde{\mathbf{E}}=\{\tilde{\mathbf{e}}_t\}_{t=1}^{T'}$ and $\tilde{\mathbf{S}}=\{\tilde{\mathbf{s}}_t\}_{t=1}^{T'}$, yielding $\mathbf{h}^{\mathrm{emo}}_t = \mathrm{CrossAttn}(\tilde{\mathbf{z}}_t, \tilde{\mathbf{E}}, \tilde{\mathbf{E}})$ and $\mathbf{h}^{\mathrm{sem}}_t = \mathrm{CrossAttn}(\tilde{\mathbf{z}}_t, \tilde{\mathbf{S}}, \tilde{\mathbf{S}})$.
The resulting modulation signals are projected back to the acoustic latent space using a shared linear mapping, producing $\mathbf{u}^{\mathrm{emo}}_t = W_m \mathbf{h}^{\mathrm{emo}}_t$ and $\mathbf{u}^{\mathrm{sem}}_t = W_m \mathbf{h}^{\mathrm{sem}}_t$. To balance emotion and semantic contributions, we apply independent stochastic dropout to each modulation term and form the unified latent representation as $\mathbf{z}^{\mathrm{uni}}_t = \mathbf{z}_t + (\mathbf{u}^{\mathrm{emo}}_t \odot \mathbf{d}^{\mathrm{emo}}_t) + (\mathbf{u}^{\mathrm{sem}}_t \odot \mathbf{d}^{\mathrm{sem}}_t)$, where $\mathbf{d}^{\mathrm{emo}}_t, \mathbf{d}^{\mathrm{sem}}_t \in \{0,1\}^{D}$ are independently sampled. 
The resulting unified latents integrate emotion-aware and semantic-aware refinements while preserving the underlying acoustic structure prior to residual vector quantization.

\subsubsection{Relation-Preserving Emotion--Semantic Distillation}
\label{stage:2}

Residual vector quantization (RVQ) can disrupt relational structure when mapping
continuous representations to discrete codes. To address this issue, we adopt a
relation-preserving distillation strategy~\cite{wang2024distilvpr} that supervises
the first-layer quantized representations by aligning their pairwise geometric
relations with those from emotion and semantic teacher spaces. Given frame-level
emotion features $\mathbf{E}=\{\mathbf{e}_t\}_{t=1}^{T'} \in \mathbb{R}^{T' \times D_{\mathrm{e}}}$
and semantic audio features $\mathbf{S}=\{\mathbf{s}_t\}_{t=1}^{T'} \in \mathbb{R}^{T' \times D_{\mathrm{s}}}$,
we define teacher relational descriptors for each timestep pair $(t,t')$ using
Euclidean distances:
$r^{\mathrm{emo}}_{t,t'} = \|\mathbf{e}_t - \mathbf{e}_{t'}\|_2$
and
$r^{\mathrm{sem}}_{t,t'} = \|\mathbf{s}_t - \mathbf{s}_{t'}\|_2$.
Prior work has shown that the first RVQ layer captures particularly informative
and structured representations~\cite{xin2024speechtokenizer}. Accordingly, we compute
student relational descriptors from the first residual quantized outputs
$\mathbf{Q}^{(1)}=\{\mathbf{Q}^{(1)}_t\}_{t=1}^{T'}$ as
$r^{(1)}_{t,t'} = \|\mathbf{Q}^{(1)}_t - \mathbf{Q}^{(1)}_{t'}\|_2$.
Relational consistency between student and teacher is enforced via
\begin{equation}
\mathcal{L}_{\mathrm{rela}} =
\frac{1}{T'^2}
\sum_{t,t'}
\Big[
\alpha \, d\!\left(r^{(1)}_{t,t'}, r^{\mathrm{emo}}_{t,t'}\right)
+
\beta \, d\!\left(r^{(1)}_{t,t'}, r^{\mathrm{sem}}_{t,t'}\right)
\Big],
\end{equation}
where $d(\cdot,\cdot)$ denotes an $\ell_1$ discrepancy. This objective preserves emotion- and semantic-relevant relational structure under discretization.

\subsubsection{Emotion-Weighted Semantic Alignment}
\label{stage:3}

To address the inherent mismatch in sequence length between frame-level RVQ outputs and token-level contextual embeddings, we perform semantic alignment between quantized representations and textual semantics. Moreover, to mitigate emotion degradation introduced by discrete quantization, we impose stronger supervision on frames exhibiting larger emotion variation, which are empirically more susceptible to affective distortion. The complete procedure is summarized in Algorithm~\ref{alg:emo_align}. 

Let the quantized latent sequence be $\mathbf{Q}^{(1)}=\{\mathbf{Q}^{(1)}_t\}_{t=1}^{T'}$, the textual semantic embeddings be $\mathbf{C}=\{\mathbf{c}_i\}_{i=1}^{n}$, and the frame-level emotion features be $\mathbf{E}=\{\mathbf{e}_t\}_{t=1}^{T'}$. Assuming a roughly monotonic correspondence between speech frames and text tokens, each frame $t$ is associated with a local textual neighborhood without dynamic programming. We compute a center index $i_0(t)=\lfloor t n / T' \rfloor$ and define a window $\mathcal{N}(t)$ of width $2w+1$. Within this window, cosine similarities between $\mathbf{Q}^{(1)}_t$ and $\mathbf{c}_i$ are normalized via softmax to obtain alignment weights $a_{t,i}$, which are used to construct a semantic teacher $\mathbf{c}_t^{*}=\sum_{i\in\mathcal{N}(t)} a_{t,i}\mathbf{c}_i$. To explicitly realize this emotion-aware supervision, \emph{Emotion Variation
} is operationalized as the magnitude of frame-level emotion variation, quantified by differences in emotion embeddings across adjacent frames. Accordingly, we compute a framewise emotion difference magnitude as $d_t=\lVert \mathbf{e}_t-\mathbf{e}_{t-1}\rVert_1$ with $d_1=0$, and derive normalized importance weights $\gamma_t$ by applying a softmax over $\{d_t\}$ and scaling to unit mean. These weights modulate each frame's contribution in the alignment objective, which is defined as
\begin{equation}
\mathcal{L}_{\mathrm{align}}
=
-\frac{1}{T'}
\sum_{t=1}^{T'}
\gamma_t \log \sigma\!\left(\cos(\mathbf{Q}^{(1)}_t,\mathbf{c}_t^{*})\right),
\tag{2}
\end{equation}
where $\sigma(\cdot)$ denotes the sigmoid function.

\begin{algorithm}[t]
\caption{Emo-Weighted Semantic Alignment}
\label{alg:emo_align}
\begin{algorithmic}[1]

\Require Quantized representations $\{\mathbf{Q}^{(1)}_t\}_{t=1}^{T'}$, 
textual semantic embeddings $\{\mathbf{c}_i\}_{i=1}^{n}$, 
emotion features $\{\mathbf{e}_t\}_{t=1}^{T'}$, 
window size $w$

\State Compute framewise emotion differences:
\Statex\hspace{\algorithmicindent}
$d_1 = 0,\quad d_t = \lVert \mathbf{e}_t - \mathbf{e}_{t-1} \rVert_1,\; t = 2,\dots,T'$

\State Compute emotion importance weights:
\Statex\hspace{\algorithmicindent}
$\boldsymbol{\gamma} = T' \cdot \mathrm{Softmax}(\{d_t\}_{t=1}^{T'})$

\For{$t = 1$ to $T'$}

    \State Compute center index:
    \Statex\hspace{\algorithmicindent}
    $i_0 = \mathrm{clip}\!\left(\left\lfloor t \cdot \frac{n}{T'} \right\rfloor,\; 1,\; n\right)$

    \State Define local neighborhood:
    \Statex\hspace{\algorithmicindent}
    $\mathcal{N}(t) = \{\, i \mid |i - i_0| \le w \,\}$

    \State Compute cosine similarities:
    \Statex\hspace{\algorithmicindent}
    $s_{t,i} = \cos(\mathbf{Q}^{(1)}_t , \mathbf{c}_i),\quad i \in \mathcal{N}(t)$

    \State Compute alignment weights:
    \Statex\hspace{\algorithmicindent}
    $a_{t,i} = \exp(s_{t,i}) \big/ \sum_{j \in \mathcal{N}(t)} \exp(s_{t,j})$

    \State Construct semantic teacher:
    \Statex\hspace{\algorithmicindent}
    $\mathbf{c}_t^{*} = \sum_{i\in\mathcal{N}(t)} a_{t,i}\, \mathbf{c}_i$

\EndFor

\State \Return $\{\mathbf{c}_t^{*}\}_{t=1}^{T'}$ and $\{\gamma_t\}_{t=1}^{T'}$

\end{algorithmic}

\end{algorithm}

\subsection{Training Objective}
We adopt a multi-objective training framework that combines standard neural codec
reconstruction losses~\cite{encodec_defossez2022high, xin2024speechtokenizer, ji2024wavtokenizer}
with two emotion–semantic supervisory objectives. The generator is optimized with four
reconstruction losses: the quantization commitment loss $\mathcal{L}_q$, the
mel-spectrogram loss $\mathcal{L}_{\mathrm{mel}}$, the adversarial loss
$\mathcal{L}_{\mathrm{adv}}$, and the feature matching loss $\mathcal{L}_{\mathrm{feat}}$.
In addition, we regularize discrete representations using two objectives:
the relation-preserving distillation loss $\mathcal{L}_{\mathrm{rela}}$, which enforces
emotion and semantic relational consistency from teacher representations, and the
emotion-weighted semantic alignment loss $\mathcal{L}_{\mathrm{align}}$, which aligns
quantized tokens with contextual semantics while emphasizing emotionally salient frames.
The overall training objective is defined as
\begin{align}
\mathcal{L}_{\mathrm{total}}
&=
\lambda_{\mathrm{mel}} \mathcal{L}_{\mathrm{mel}}
+ \lambda_{\mathrm{adv}} \mathcal{L}_{\mathrm{adv}}
+ \lambda_{\mathrm{feat}} \mathcal{L}_{\mathrm{feat}}
\nonumber\\
&\quad
+ \lambda_q \mathcal{L}_q + \lambda_{\mathrm{rela}} \mathcal{L}_{\mathrm{rela}}
+ \lambda_{\mathrm{align}} \mathcal{L}_{\mathrm{align}} .
\tag{3}
\end{align}
Detailed training objectives are in Appendix~\ref{apx:training_objective}.

\subsection{Downstream Extension to TTS}
The emotion-aware discrete representations learned by the codec are further utilized for text-to-speech synthesis by training a language model over RVQ tokens, following prior work~\cite{xin2024speechtokenizer,zero_tts_wang2023neural}. Conditioned on a phoneme sequence $\mathbf{y}$ and a reference acoustic prompt $\mathbf{P} \in \mathbb{R}^{T' \times K}$, the model generates $K$ parallel token streams aligned with the RVQ hierarchy. The first stream, which encodes coarse linguistic structure and global prosodic, is modeled autoregressively using a decoder-only Transformer, optimized with
\begin{equation}
\mathcal{L}_{\mathrm{AR}}
=
- \log \prod_{t=1}^{T'}
p\!\left(u^{(1)}_{t} \;\middle|\; u^{(1)}_{<t},\, \mathbf{y};\, \theta_{\mathrm{AR}}\right).
\tag{4}
\end{equation}
Subsequent RVQ streams encode finer acoustic detail and emotional variation.
For layers $k = 2,\dots,K$, a non-autoregressive Transformer infers the entire
token sequence $\mathbf{u}^{(k)}$ conditioned on previously inferred layers,
the phoneme sequence, and the acoustic prompt:
\begin{equation}
\mathcal{L}_{\mathrm{NAR}}
=
- \log \prod_{k=2}^{K}
p\!\left(\mathbf{u}^{(k)} \;\middle|\; \mathbf{u}^{(<k)},\, \mathbf{y},\, \mathbf{P};\, \theta_{\mathrm{NAR}}\right).
\tag{5}
\end{equation}
Both components share an identical Transformer configuration with 12 layers,
16 attention heads, 1024-dimensional embeddings, 4096-dimensional feed-forward
layers, and a dropout rate of 0.05. The inferred RVQ token hierarchies are
subsequently mapped to discrete embeddings $\mathbf{Q}^{(k)}$ and decoded to
synthesize speech under joint textual, acoustic, and emotion-aware conditioning.

\section{Experiments}
\begin{table}[t]
\centering
\scriptsize
\begin{tabular}{l c c c c c}
\toprule
\textbf{Model} & \textbf{BR$\downarrow$} & \textbf{FR$\downarrow$} & \textbf{Nq$\downarrow$} & \textbf{Train. Data$\downarrow$} & \textbf{Param$\downarrow$} \\
\midrule
Encodec         & 6   & 75   & 8  & 17 & 20  \\
DAC             & 6  & 50   & 12  & 8  & 76 \\
FACodec         & 4.8   & 80   & 6  & 500 & 500 \\
BigCodec        & 1  & 80   & 1  & 1  & 159 \\
Mimi            & 1.1   & 12.5 & 8  & 1  & 95 \\
TAAE            & 0.6 & 25   & 1  & 100  & 950 \\
WavTokenizer    & 0.9   & 75   & 1  & 80  & 40\\
SpeechTokenizer & 4     & 50   & 8  & 1 & 18 \\
Llasa           & 0.8   & 50   & 1  &  150 &  1000\\
\midrule
\textbf{Ours}   & 4     & 50   & 8  & 2.3  & 44 \\
\bottomrule
\end{tabular}
\caption{Comparison of codecs across efficiency dimensions. 
BR: Bitrate (kbps); FR: Frame Rate (Hz); Nq: Number of Quantizers; Train. Data: Trainng Hours (k hours); 
Param: Number of Parameters (M).}
\label{tab:motivation-efficiency}
\vspace{-4mm}
\end{table}

\subsection{Experiment Setups}
\begin{table*}[t]
\centering
\caption{
\textbf{Objective speech reconstruction results} are reported across emotional consistency, information  preservation, and speech naturalness, evaluated on EmoVoiceDB and the LibriSpeech test-clean set.
\textbf{Bold} marks best scores, and \underline{underline} indicates second-best scores. Results are averaged over three random seeds.
}
\label{tab:recon_evaluation_clean}
\resizebox{0.9\textwidth}{!}{
\begin{tabular}{l ccc ccc ccc}
\toprule
\multirow{2}{*}{\textbf{Model}} 
& \multicolumn{3}{c}{\textbf{Emotional Consistency
}} 
& \multicolumn{3}{c}{\textbf{Information Preservation}} 
& \multicolumn{3}{c}{\textbf{Speech Naturalness}} \\
\cmidrule(lr){2-4} \cmidrule(lr){5-7} \cmidrule(lr){8-10}
& \textbf{Emo SIM$\uparrow$} & \textbf{Pros SIM$\uparrow$} & \textbf{Recall$\uparrow$}
& \textbf{WER$\downarrow$} & \textbf{WIL$\downarrow$} & \textbf{LSD$\downarrow$}
& \textbf{MSEP$\downarrow$} & \textbf{PESQ$\uparrow$} & \textbf{UTMOS$\uparrow$} \\
\midrule
EnCodec 
& \cellcolor{blockgreen}0.73 & \cellcolor{blockgreen}0.78 & \cellcolor{blockgreen}0.37
& \cellcolor{blockyellow}\textbf{4.02} & \cellcolor{blockyellow}6.63 & \cellcolor{blockyellow}0.97
& \cellcolor{blockblue}35.06 & \cellcolor{blockblue}2.38 & \cellcolor{blockblue}2.43 \\

DAC 
& \cellcolor{blockgreen}0.79 & \cellcolor{blockgreen}0.74 & \cellcolor{blockgreen}0.31
& \cellcolor{blockyellow}\underline{4.10} & \cellcolor{blockyellow}\underline{6.52} & \cellcolor{blockyellow}0.94
& \cellcolor{blockblue}30.36 & \cellcolor{blockblue}2.74 & \cellcolor{blockblue}3.31 \\

FACodec 
& \cellcolor{blockgreen}\underline{0.88}& \cellcolor{blockgreen}0.70 & \cellcolor{blockgreen}0.32
& \cellcolor{blockyellow}4.14 & \cellcolor{blockyellow}6.64 & \cellcolor{blockyellow}0.85
& \cellcolor{blockblue}\textbf{12.16} & \cellcolor{blockblue}\underline{2.85} & \cellcolor{blockblue}3.49 \\

SpeechTokenizer 
& \cellcolor{blockgreen}0.82 & \cellcolor{blockgreen}0.77 & \cellcolor{blockgreen}0.29
& \cellcolor{blockyellow}4.19 & \cellcolor{blockyellow}6.78 & \cellcolor{blockyellow}1.07
& \cellcolor{blockblue}40.38 & \cellcolor{blockblue}2.58 & \cellcolor{blockblue}3.43 \\

Mimi 
& \cellcolor{blockgreen}0.85 & \cellcolor{blockgreen}0.78 & \cellcolor{blockgreen}0.35
& \cellcolor{blockyellow}10.72 & \cellcolor{blockyellow}16.34 & \cellcolor{blockyellow}1.08
& \cellcolor{blockblue}43.24 & \cellcolor{blockblue}1.65 & \cellcolor{blockblue}2.29 \\

BigCodec 
& \cellcolor{blockgreen}0.78 & \cellcolor{blockgreen}0.71 & \cellcolor{blockgreen}0.32
& \cellcolor{blockyellow}4.58 & \cellcolor{blockyellow}7.45 & \cellcolor{blockyellow}\underline{0.84}
& \cellcolor{blockblue}22.91 & \cellcolor{blockblue}2.68 & \cellcolor{blockblue}3.44 \\

TAAE 
& \cellcolor{blockgreen}0.84 & \cellcolor{blockgreen}0.73 & \cellcolor{blockgreen}0.33
& \cellcolor{blockyellow}9.35 & \cellcolor{blockyellow}13.81 & \cellcolor{blockyellow}1.05
& \cellcolor{blockblue}42.35 & \cellcolor{blockblue}2.03 & \cellcolor{blockblue}\underline{3.57} \\

% TAAE 
% & \cellcolor{blockgreen}0.91 & \cellcolor{blockgreen}\underline{0.94} & \cellcolor{blockgreen}0.83
% & \cellcolor{blockyellow}13.28 & \cellcolor{blockyellow}10.11 & \cellcolor{blockyellow}1.30
% & \cellcolor{blockblue}28.59 & \cellcolor{blockblue}2.13 & \cellcolor{blockblue}3.36 \\

WavTokenizer 
& \cellcolor{blockgreen}0.83 & \cellcolor{blockgreen}\underline{0.81} & \cellcolor{blockgreen}0.38
& \cellcolor{blockyellow}6.22 & \cellcolor{blockyellow}9.16 & \cellcolor{blockyellow}0.96
& \cellcolor{blockblue}39.35 & \cellcolor{blockblue}2.19 & \cellcolor{blockblue}3.36 \\

Llasa 
& \cellcolor{blockgreen}0.87 & \cellcolor{blockgreen}0.80 & \cellcolor{blockgreen}\underline{0.40}
& \cellcolor{blockyellow}4.46 & \cellcolor{blockyellow}7.20 & \cellcolor{blockyellow}0.92
& \cellcolor{blockblue}27.49 & \cellcolor{blockblue}2.43 & \cellcolor{blockblue}3.55 \\

\midrule
\modelname 
& \cellcolor{blockgreen}\textbf{0.94} & \cellcolor{blockgreen}\textbf{0.86} & \cellcolor{blockgreen}\textbf{0.48}
& \cellcolor{blockyellow}4.15 & \cellcolor{blockyellow}\textbf{6.43} & \cellcolor{blockyellow}\textbf{0.78}
& \cellcolor{blockblue}\underline{19.21} & \cellcolor{blockblue}\textbf{3.04} & \cellcolor{blockblue}\textbf{3.68} \\
\bottomrule
\end{tabular}
}
\end{table*}
\textbf{Datasets.}
We train our model on a multi-domain speech corpus of approximately 2.3K hours,
covering three complementary aspects: clean read speech for high-fidelity
reconstruction, multilingual and acoustically diverse speech for robustness,
and emotionally expressive speech for affective representation learning.
Specifically, LibriSpeech~\cite{panayotov2015librispeech} and
VCTK~\cite{VCTK_yamagishi2019cstr} are used as clean English read-speech sources
for reconstruction, AISHELL-3~\cite{shi2020aishell} provides Mandarin speech for
multilingual generalization, and a subset of AudioSet~\cite{gemmeke2017audio}
introduces broad acoustic variability. To expose the codec to natural emotional
expressions, we additionally incorporate emotionally annotated conversational
speech from MSP-Podcast~\cite{busso2025msp} and
CMU-MOSEI~\cite{CMU_MOSEI_bagher-zadeh-etal-2018-multimodal}. All audio is
resampled to 16~kHz. For evaluation, speech reconstruction quality is measured on the LibriSpeech
test-clean set~\cite{panayotov2015librispeech}. To assess emotion
preservation under codec resynthesis, we follow the EMO-SUPERB~\cite{wu2024_emo_superb} protocol and
evaluate speech emotion recognition on synthesized speech using the official
EMO-SUPERB test partitions. For downstream
text-to-speech generation, the token prediction module is trained on
LibriTTS~\cite{zen2019libritts}, and evaluation covers linguistic intelligibility
on LibriSpeech test-clean and emotion preservation on
EmoVoiceDB~\cite{wu2024_emo_superb} and SECAP~\cite{xu2024secap}. Detailed dataset are in the Appendix~\ref{apx:datasets_detail}.

\noindent\textbf{Architecture. }
The speech tokenizer comprises an acoustic encoder, a residual vector
quantization (RVQ) module with eight quantization layers and codebooks of size
1024, and a decoder $\operatorname{Dec}(\cdot)$, following the setting in~\cite{xin2024speechtokenizer}. Adversarial training employs three discriminators,
including multi-period, multi-scale, and multi-scale STFT discriminators.
Quantization is performed at 50~Hz, with both the encoder and RVQ using an embedding dimension of $D=1024$. To provide affective and semantic guidance, we incorporate several frozen
pre-trained encoders, including CLAP-LAION (630k-best)~\cite{clap_wu2023large} as the emotion encoder, wav2vec~2.0 (base-960h)~\cite{baevski2020wav2vec} as the ASR model, BERT (bert-base)~\cite{devlin2019bert}
as the language encoder, and HuBERT (base-ls960)~\cite{hsu2021hubert} as the self-supervised speech
model. All pre-trained encoders output embeddings with $D_{\mathrm{e}} = D_{\mathrm{s}} = D_{\mathrm{c}} = 768$. Cross-attention modules are implemented with eight attention heads. Additional details are in the Appendix~\ref{apx:model_details}.

% \noindent\textbf{Baselines.}
% We compare our method against a diverse set of state-of-the-art speech tokenizers
% and neural codecs, including EnCodec~\cite{encodec_defossez2022high},
% DAC~\cite{dac_kumar2023high}, FACodec~\cite{fac_ju2024naturalspeech},
% SpeechTokenizer~\cite{xin2024speechtokenizer}, Mimi~\cite{mimi_defossez2024moshi},
% BigCodec~\cite{xin2024bigcodec}, TAAE~\cite{parker2024scaling},
% WavTokenizer~\cite{ji2024wavtokenizer}, and Llasa~\cite{ye2025llasa}. All baselines
% are evaluated using their official released checkpoints and default
% configurations; details are provided in \textbf{Appendix~C}.

\noindent\textbf{Implementation Details and metrics.}
Our codec model is trained for 200 epochs on four A100 GPUs with a batch size of
16, using the AdamW optimizer with a learning rate of $2\times10^{-4}$ and the learning rate is
decayed based on a cosine schedule. We set $\alpha = \beta = 1$ for distillation. The downstream TTS models are trained on four
A100 GPUs using ScaledAdam with a learning rate of $5\times10^{-2}$ and 120
warm-up steps, where the AR model is trained for 300 epochs and the NAR model
for 200 epochs. Training employs dynamic batching, with each batch containing
up to 550 seconds of audio for AR and 100--200 seconds for NAR.

We evaluate our model along three dimensions: emotional consistency, content preservation, and speech naturalness.
Emotional consistency is assessed using Emotion Similarity computed from emotion2vec embeddings~\cite{ma2024emotion2vec}, Prosody Similarity derived from pitch, energy, and duration features via AutoPCP~\cite{autopcp_barrault2023seamless,wu2024laugh}, and emotion recognition recall measured by a pretrained SER model~\cite{ma2024emotion2vec}.
Content preservation is evaluated through Whisper-based transcription accuracy, including Word Error Rate (WER), Word Information Lost (WIL)\cite{WIL_morris2004and}, and Log-Spectral Distance (LSD).
Speech naturalness is measured using pitch reconstruction error (MSEP), PESQ~\cite{rix2002perceptual} for perceptual quality under signal distortion, and UTMOS~\cite{chen2022wavlm}, which predicts human-judged speech naturalness.

\subsection{Main Results}

\subsubsection{Evaluation on Reconstruction}
\begin{table*}[ht]
\centering
\vspace{-4mm}
\caption{
\textbf{Speech emotion recognition performance} on the \texttt{EMO-SUPERB} benchmark.
Macro-F1 scores are reported for six emotion-focused evaluation sets.
\textbf{Bold} indicates the best result, and \underline{underline} marks the second-best.
Results are averaged over three random seeds.
}
\label{tab:ser_codec}
\resizebox{0.92\textwidth}{!}{
\begin{tabular}{l cc | cccccc}
\toprule
\multirow{2}{*}{\textbf{Model}} & \multicolumn{2}{c|}{\textbf{Codec Information}} &
\multicolumn{6}{c}{\textbf{Speech Emotion Recognition (Macro-F1) $\uparrow$}} \\
\cmidrule(lr){2-3} \cmidrule(lr){4-9}
 & \textbf{kbps} & \textbf{Configuration} &
\textbf{IEMOCAP} &
\textbf{CREMA-D} &
\textbf{IMPROV} &
\textbf{PODCAST} &
\textbf{NNIME} &
\textbf{BIIC-POD.} \\
\midrule
Original Audio & - & - &
0.313 & 0.594 & 0.491 & 0.301 & 0.183 & 0.247 \\
\midrule
AudioDec        & 6.4 & symAD\_libritts\_24000\_hop300 &
0.301 & 0.548 & 0.461 & 0.298 & 0.180 & 0.242 \\
AcademiCodec    & 2   & large universal &
0.301 & 0.548 & 0.461 & 0.298 & 0.181 & 0.242 \\
SpeechTokenizer & 4   & hubert\_avg &
0.305 & 0.573 & 0.448 & 0.292 & 0.180 & 0.243 \\
DAC             & 6   & DAC\_16k &
\underline{0.315} & \underline{0.591} & \underline{0.491} & 0.302 & \textbf{0.184} & \underline{0.247} \\
EnCodec & 1.5 & 24k &
0.280 & 0.411 & 0.321 & 0.262 & 0.166 & 0.227 \\
EnCodec & 3   & 24k &
0.275 & 0.457 & 0.448 & 0.293 & 0.178 & 0.240 \\
EnCodec & 6   & 24k &
0.295 & 0.499 & 0.450 & 0.294 & 0.178 & 0.239 \\
FunCodec & 8 & en\_libritts\_16k\_nq32ds640 &
0.312 & 0.569 & 0.482 & \underline{0.303} & 0.181 & 0.246 \\
FunCodec & 8 & zh\_en\_16k\_nq32ds640 &
0.312 & 0.577 & 0.482 & 0.302 & 0.182 & 0.246 \\
Soundstream & 6 & Soundstream &
0.261 & 0.411 & 0.321 & 0.262 & 0.146 & 0.213 \\
MP3 & 6 & - &
0.259 & 0.405 & 0.321 & 0.262 & 0.149 & 0.210 \\
Opus & 6 & - &
0.271 & 0.433 & 0.338 & 0.269 & 0.162 & 0.226 \\
AAC & 6 & - &
0.268 & 0.428 & 0.334 & 0.268 & 0.165 & 0.227 \\
\midrule
\textbf{Ours} & 4 & 16k &
\textbf{0.338} & \textbf{0.629} & \textbf{0.513} & \textbf{0.319} & \underline{0.182} & \textbf{0.256} \\
\bottomrule
\end{tabular}
}
\vspace{-4mm}
\end{table*}

We evaluate reconstruction performance on two benchmarks, as shown in Table~\ref{tab:recon_evaluation_clean}: EmoVoiceDB, which assesses emotion preservation in reconstructed speech, and LibriSpeech test-clean, which measures intelligibility and perceptual quality following prior work~\cite{tts_ji2024mobilespeech,xin2024bigcodec}. Complete baseline details are in the Appendix~\ref{apx:baseline}.

\noindent\textbf{Emotional Consistency.}
Our method achieves the strongest emotion preservation across all metrics, with Emo SIM of 0.94 surpassing FACodec (0.88) and Pros SIM of 0.86 exceeding WavTokenizer~\cite{ji2024wavtokenizer} (0.81). SER-based emotion recall reaches 0.48, notably higher than Llasa (0.40), indicating improved preservation of both global emotion and fine-grained prosody.

\noindent\textbf{Information Preservation.}
On LibriSpeech, the proposed codec maintains competitive intelligibility while explicitly optimizing emotion preservation. Although EnCodec achieves the lowest WER (4.02), our method remains comparable (4.15) and yields stronger complementary content metrics, including lower WIL (6.43) and the lowest LSD (0.78). These results indicate that emotion-aware modeling preserves spectral detail and linguistic content without sacrificing expressiveness.

\noindent\textbf{Speech Naturalness.}
Our codec achieves the best perceptual quality, with the highest PESQ (3.04) and UTMOS (3.68). Although MSEP is not the lowest, improvements in perceptual metrics indicate cleaner and more natural reconstructions. Our codec better balances emotional fidelity, content preservation, and perceptual naturalness than prior approaches. Subjective results and representation quality evaluation are in the Appendix~\ref{apx:subjective_reconstruction} and \ref{apx:subjective_tts}.

\subsubsection{Speech Emotion Recognition Evaluation}
\begin{table*}[t]
\caption{\textbf{TTS Evaluation Results.} We evaluate zero-shot TTS performance across three datasets: LibriSpeech, EmoVoiceDB, and SECAP. Each evaluation reports content preservation (WER), speaker similarity (SIM), emotional alignment (Emo\_SIM), and naturalness (UTMOS). Results are averaged over three random seeds.}
\label{tab:tts-results}
\begin{center}
\vspace{-6mm}
\resizebox{\textwidth}{!}{
\begin{threeparttable}
\begin{tabular}{l|c|cccc|cccc|cccc}
\toprule
\multirow{3}{*}{\textbf{System}} & \multirow{2}{*}{\textbf{Frame}} 
& \multicolumn{4}{c|}{\textbf{LibriSpeech}} 
& \multicolumn{4}{c|}{\textbf{EmoVoice-DB}} 
& \multicolumn{4}{c}{\textbf{SECAP}} \\
\cmidrule(lr){3-14}
 & \textbf{Rate} 
 & \textbf{WER$\downarrow$} & \textbf{SIM-O$\uparrow$} & \textbf{WIL$\downarrow$} & \textbf{UTMOS$\uparrow$}
 & \textbf{WER$\downarrow$} & \textbf{Emo\_SIM$\uparrow$} & \textbf{Recall$\uparrow$} & \textbf{UTMOS$\uparrow$}
 & \textbf{WER$\downarrow$} & \textbf{Emo\_SIM$\uparrow$} & \textbf{Recall$\uparrow$} & \textbf{UTMOS$\uparrow$} \\
\midrule
\rowcolor{green!8} \multicolumn{14}{l}{\textbf{\textit{NAR Models}}} \\
MaskGCT & 50 
& 2.63 & 0.68 & 13.83 & 3.10 
& 3.28 & 0.71 & 0.33 & 3.25
& 9.52 & 0.74 & 0.36 & 2.70 \\
F5-TTS & 93.75
& 2.53 & 0.66 & 11.10 & 3.25
& 3.45 & 0.69 & 0.31 & 3.30
& 8.87 & 0.72 & 0.34 & 2.65 \\
\midrule

\rowcolor{blue!5} \multicolumn{14}{l}{\textbf{\textit{AR Models}}} \\
FireRedTTS & 25
& 2.69 & 0.47 & 15.47 & 3.05
& 3.61 & 0.57 & 0.29 & 3.10
& 9.13 & 0.59 & 0.31 & 2.75 \\

ARS & 50
& 2.64 & 0.68 & 10.88 & 3.45
& 3.42 & 0.74 & 0.34 & 3.50
& 8.65 & 0.72 & 0.38 & 2.95 \\

CosyVoice 2 & 25
& \textbf{2.45} & 0.77 & \textbf{6.72} & 4.23
& 3.47 & 0.87 & 0.37 & \textbf{4.42}
& \textbf{8.55} & 0.79 & 0.43 & 2.75 \\

Llasa & 50
& 2.49 & 0.58 & 9.34 & 3.55
& 3.61 & 0.70 & 0.32 & 3.40
& 8.92 & 0.69 & 0.35 & 2.85 \\

SparkTTS & 50
& 2.57 & 0.78 & 7.18 & 4.17
& \textbf{3.28} & 0.82 & 0.36 & 4.20
& 9.03 & 0.77 & 0.40 & 2.95 \\

\midrule
\textbf{Ours} & 50
& 2.51 & \textbf{0.80} & 7.29 & \textbf{4.29}
& 3.48 & \textbf{0.91} & \textbf{0.41} & 4.35
& 8.64 & \textbf{0.84} & \textbf{0.49} & \textbf{3.20} \\
\bottomrule
\end{tabular}
\end{threeparttable}
}
\end{center}
\vspace{-5mm}
\end{table*}
\begin{table*}[t]
\centering
\small
\resizebox{\textwidth}{!}{
\begin{tabular}{c c c | c c c c | c c c c}
\toprule
\multicolumn{3}{c|}{\textbf{Model Components}} &
\multicolumn{4}{c|}{\textbf{Reconstruction}} &
\multicolumn{4}{c}{\textbf{TTS}} \\
\midrule
\textbf{EG-Latent} & \textbf{RP-Distill} & \textbf{EW-Align} &
\textbf{Emo SIM} $\uparrow$ &
\textbf{WER} $\downarrow$ &
\textbf{UTMOS} $\uparrow$ &
\textbf{MUSHRA} $\uparrow$ &
\textbf{Emo SIM} $\uparrow$ &
\textbf{WER} $\downarrow$ &
\textbf{UTMOS} $\uparrow$ &
\textbf{MOS} $\uparrow$ \\
\midrule
  &   &   &
0.87 & 4.85 & 3.34 & 86.27 &
0.86 & 4.12 & 3.85 & 3.68 \\
\checkmark &   &   &
0.90 & 4.62 & 3.46 & 88.31 &
0.88 & 3.95 & 3.98 & 3.80 \\
  & \checkmark &   &
0.89 & 4.47 & 3.41 & 87.54 &
0.89 & 3.69 & 4.01 & 3.77 \\
  &   & \checkmark &
0.88 & 4.53 & 3.49 & 87.12 &
0.87 & 3.81 & 4.06 & 3.83 \\
\midrule
\checkmark & \checkmark &   &
0.93 & 4.30 & 3.54 & 89.63 &
0.90 & 3.55 & 4.15 & 3.92 \\
\checkmark &   & \checkmark &
0.92 & 4.42 & 3.57 & 89.94 &
0.89 & 3.62 & 4.22 & 3.98 \\
  & \checkmark & \checkmark &
0.90 & 4.21 & 3.52 & 88.86 &
0.90 & 3.51 & 4.18 & 3.90 \\
\midrule
\checkmark & \checkmark & \checkmark &
\textbf{0.94} & \textbf{4.15} & \textbf{3.68} & \textbf{90.71} &
\textbf{0.91} & \textbf{3.48} & \textbf{4.35} & \textbf{4.05} \\
\bottomrule
\end{tabular}
}
\caption{
Ablation study on the three components: \textbf{E}motion-\textbf{S}emantic \textbf{G}uided \textbf{Latent} (EG-Latent), \textbf{R}elation-\textbf{P}reserving \textbf{Distill}ation (RP-Distill), and \textbf{E}motion-\textbf{W}eighted Soft Semantic \textbf{Align}ment (EW-Align). We assess emotional expressiveness (Emo SIM), semantic preservation (WER), prosodic naturalness (UTMOS), and subjective perceptual quality (MUSHRA/MOS) for both reconstruction and TTS. Following prior settings, LibriSpeech evaluates semantic fidelity and prosodic naturalness, while EmoVoice-DB assesses emotion preservation.
}
\label{tab:ablation_rec_tts}
\vspace{-5.5mm}
\end{table*}

Reconstruction-based metrics do not directly reflect whether emotion-related
information remains usable for downstream perception. We therefore evaluate our
codec on the EMO-SUPERB benchmark, which measures emotion discrimination
performance across six standard SER datasets using macro-F1. Results are
reported in Table~\ref{tab:ser_codec}.

\noindent\textbf{Trend across English datasets.}
Across IEMOCAP~\cite{busso2008iemocap}, CREMA-D~\cite{cao2014crema}, and IMPRoV~\cite{busso2016msp}, our model consistently ranks first or second
among all codec baselines, indicating strong emotion retention. Notably, on
IMPRoV, our representation even outperforms original audio, suggesting that the
learned discrete space suppresses nuisance variability such as channel mismatch
and recording artifacts, which benefits downstream emotion classification.

\noindent\textbf{Variability on Chinese datasets.}
On BIIC-PODCAST~\cite{BIIC_upadhyay2023intelligent}, our codec achieves the best F1 score, showing robustness to conversational speech and diverse recording conditions. Performance on NNIME~\cite{chou2017nnime} is slightly below the strongest baseline, due to its subtle and regulated emotional expressions, which remain challenging for discrete representations.

\noindent\textbf{General observations.} Our codec generalizes across languages, outperforming neural and legacy codecs at comparable bitrates and approaching original emotion discriminability.

\subsubsection{Zero-shot TTS Generation Evaluation}

To assess whether the learned codec representations generalize beyond reconstruction, we evaluate their effectiveness in zero-shot text-to-speech (TTS) synthesis (Table~\ref{tab:tts-results}). Following prior work, the TTS model is trained on LibriTTS and evaluated on three benchmarks: LibriSpeech for intelligibility and prosodic naturalness (WER, WIL, SIM-O~\cite{x-codec_ye2025codec}, UTMOS), and EmoVoice-DB and SECAP for emotion alignment (Emo SIM, Recall) under expressive speech. Our goal is not to optimize TTS quality itself, but to examine whether the codec representations support intelligible, prosodically coherent, and emotionally aligned synthesis. Subjective results are provided in Appendix~\ref{apx:subjective_tts}.

\noindent\textbf{Linguistic precision and prosodic fluency.}
On LibriSpeech, our codec achieves the highest prosody similarity (SIM-O = 0.80) with competitive naturalness (UTMOS = 4.29) and strong intelligibility (WER = 2.51). Although CosyVoice-2~\cite{du2024cosyvoice} attains slightly lower WER, it shows weaker prosodic and perceptual quality. On the emotion-rich EmoVoice-DB and SECAP datasets, our method achieves the highest UTMOS scores, indicating stable expressive pitch and rhythm.

\noindent\textbf{Emotion discrimination during synthesis.}
Our model also preserves emotional intent during synthesis. It achieves the
highest Emo\_SIM and recall on both EmoVoice-DB and SECAP, confirming that the
emotion-aware cues encoded in the latent space remain discriminative after
generation. Overall, these results show that our codec enables zero-shot speech
synthesis with improved affective realism and robustness in emotionally dynamic
conditions. We release our demo publicly available.\footnote{\url{https://jiachengqaq.github.io/affectcodec_demo/}}

\subsection{Ablation}
\noindent\textbf{Effect of EG-Latent.}
As shown in Table~\ref{tab:ablation_rec_tts}, introducing EG-Latent substantially enhances emotional expressiveness and reconstruction fidelity. Specifically, Emo SIM increases from 0.87 to 0.90, while MUSHRA~\cite{encodec_defossez2022high} improves from 86.27 to 88.31. These indicate that emotion-guided latent modeling enables the codec to better preserve affective nuances during reconstruction.

\noindent\textbf{Effect of RP-Distill.}
RP-Distill aligns representations with semantic knowledge from a teacher, primarily improving linguistic clarity. Compared to the baseline, it reduces WER from 5.75 to 5.18 and slightly increases TTS Emo SIM from 0.86 to 0.89, indicating improved intelligibility while preserving emotional content for downstream generation.

\noindent\textbf{Effect of EW-Align.}
EW-Align strengthens prosodic and semantic expression, particularly at emotionally salient regions.
It increases reconstruction UTMOS from 3.34 to 3.49 and TTS Emo SIM from 0.86 to 0.87, highlighting gains in naturalness and emotion coherence. Overall, the components provide complementary improvements across emotion, content, and prosody, with their combination achieving the best performance across all metrics. Qualitative analysis and fine-grained ablations are provided in Appendix~\ref{apx:qualitative} and Appendix~\ref{apx:fine-grained ablation}.
%\vspace{-1mm}
\section{Conclusion}
We present AffectCodec, a neural speech codec that explicitly targets emotion preservation in discrete speech representations. By integrating emotion-aware latent, relation-preserving distillation, and emotion-weighted semantic alignment, the proposed codec improves emotional expressiveness while retaining semantic fidelity and prosodic naturalness. Extensive evaluations on speech reconstruction, speech emotion recognition, and zero-shot TTS synthesis demonstrate consistent gains in emotional consistency and perceptual quality across diverse speaking styles and conditions.

\section{Limitations}
The proposed codec achieves strong performance on emotion-related speech reconstruction benchmarks, demonstrating effective preservation of affective cues alongside semantic and prosodic fidelity. The framework is designed to reconstruct emotional expressiveness under acceptable computational efficiency, rather than to minimize model complexity. Future work may explore lighter-weight architectures and more efficient training strategies to further improve scalability while retaining emotion-preserving capabilities.

\section{Ethics Statement}

All speech datasets used in this work, as detailed in Appendix~\ref{apx:datasets_detail}, are publicly released for academic research purposes. We
strictly adhere to the licenses and usage terms associated with each dataset. The data do not contain
personally identifiable information (PII), and no attempt is made to identify or infer individual
identities from the speech signals. The proposed method is developed and evaluated solely in an offline research setting. No real-world
deployment or user-facing application is involved in this work.

\bibliography{custom}
\clearpage
\appendix
%\section{Appendix}
\centerline{\Large \bf Appendix}
\vspace{0.5em}
\section{Related Work}
\subsection{Neural Speech Codecs and Discrete Audio Representation}
\label{apx:related_neural_codec}
Neural audio codecs based on vector quantization~\cite{vector_q_van2017neural} have become a cornerstone of modern speech and audio generation systems. Early work such as VQ-VAE~\cite{vq-vae-van2017neural} and its residual variants introduced discrete latent representations for efficient compression and autoregressive modeling. Building on this paradigm, SoundStream~\cite{zeghidour2021soundstream} and EnCodec~\cite{encodec_defossez2022high} employed residual vector quantization with adversarial training to achieve high-fidelity reconstruction at low bitrates, while HiFi-Codec~\cite{yang2023hifi} and DAC~\cite{dac_kumar2023high} further improved efficiency and stability through group quantization and refined codebook learning. These codecs primarily optimize acoustic reconstruction quality and compression efficiency, treating emotion as an implicit attribute of the signal.

More recent work has explored codec designs tailored for language-model-based speech generation. SpeechTokenizer~\cite{xin2024speechtokenizer}, FACodec~\cite{fac_ju2024naturalspeech}, and Mimi~\cite{mimi_defossez2024moshi} introduce hierarchical or disentangled representations, often supervising semantic information in early quantization layers using ASR or self-supervised speech models. Single-codebook designs such as BigCodec~\cite{xin2024bigcodec}, WavTokenizer~\cite{ji2024wavtokenizer}, and TAAE~\cite{parker2024scaling} improve compatibility with large language models by flattening token streams, but may limit expressiveness at low token rates. Semantic-enhanced codecs such as X-Codec~\cite{x-codec_ye2025codec} incorporate pretrained speech representations into the quantization process to improve downstream modeling, while Llasa~\cite{ye2025llasa} aligns speech tokenization with standard LLM architectures using a simplified codec and Transformer framework. Despite these advances, emotional expressiveness is typically not an explicit optimization target and remains weakly constrained.

EmoCodec~\cite{emocodec_ren2024emo} highlights this limitation by systematically evaluating emotional degradation in existing codecs, showing that affective cues are often poorly preserved after quantization~\cite{codec_superb_wu2024codec}. However, emotion is primarily treated as an evaluation dimension rather than a modeling objective. In contrast, our work explicitly addresses emotion preservation in discrete speech representations by integrating emotion-aware optimization at the latent, relational, and alignment levels.

\subsection{Emotion-Aware Speech Representation Learning}
\label{apx:emotion_related_work}
Emotion-aware speech representation learning has been widely studied in speech emotion recognition and expressive speech modeling. Early approaches~\cite{hand_schuller2009interspeech,hand_ververidis2006emotional} relied on handcrafted acoustic features and prosodic descriptors, while later work adopted deep neural networks to learn emotion-discriminative representations directly from waveforms or spectrograms~\cite{envsdd,xiao2024configurable}. More recently, self-supervised speech models such as wav2vec 2.0~\cite{baevski2020wav2vec},
HuBERT~\cite{hsu2021hubert}, and WavLM~\cite{chen2022wavlm} have been shown to encode rich
affective information and transfer effectively to emotion-related tasks. In parallel,
cross-modal contrastive models such as AudioCLIP~\cite{guzhov2022audioclip}, CLAP-MST~\cite{elizalde2023clap} and CLAP-LAION~\cite{clap_wu2023large}
learn emotion-relevant representations through large-scale audio–text alignment. Building on these representations, several methods~\cite{ma2024emotion2vec,wang2024blsp,xiao2025dg,shi2025clep} further adapt or fine-tune pretrained encoders to emphasize emotional attributes, yielding strong performance on standard SER benchmarks.

Beyond recognition, emotion representations have also been incorporated into speech generation systems, particularly emotional and expressive text-to-speech models. Prior work~\cite{wu2019end,guo2023emodiff,shi2025emotion,gao2025emo} explores disentangling emotion from linguistic content and speaker identity to improve controllability and generalization. However, maintaining stable emotion expression becomes increasingly challenging in complex generation pipelines, especially those involving discrete tokenization and multi-stage synthesis, as emotional cues are often attenuated or distorted during intermediate representation transformations.

Despite these advances, existing emotion-aware representations are typically learned independently of neural speech codecs. In codec-based pipelines~\cite{encodec_defossez2022high,xin2024speechtokenizer}, emotion is commonly treated as an implicit attribute preserved through acoustic reconstruction, without explicit constraints to protect affective structure during discretization. Recent analysis-oriented work~\cite{codec_superb_wu2024codec}, such as EmoCodec~\cite{emocodec_ren2024emo}, highlights this limitation by systematically evaluating emotion degradation across codecs, but does not modify the codec learning objective itself. This gap motivates approaches that explicitly integrate emotion-aware objectives into discrete speech representation learning, bridging emotion modeling and neural codecs at the representation level.
\section{Baseline }
\label{apx:baseline}
\subsection{Speech Tokenizers and Neural Codecs}

\noindent\textbf{EnCodec.}
EnCodec~\cite{encodec_defossez2022high} is an RVQ-based neural audio codec that discretizes speech at a relatively high temporal resolution. It operates at a 75 Hz frame rate and uses two residual codebooks during inference, resulting in a bitrate of approximately 6 kbps. The model is trained with adversarial objectives using multi-scale and multi-period discriminators, and we evaluate it using the official pretrained checkpoint.

\noindent\textbf{SoundStream.}
SoundStream~\cite{zeghidour2021soundstream} is an end-to-end neural audio codec operating on 24 kHz waveform inputs, using residual vector quantization and adversarial training to get high perceptual quality at low to medium bitrates (3–18 kbps). Quantizer dropout enables bitrate scalability within a single model for real-time streaming.

\noindent\textbf{FunCodec.}
FunCodec~\cite{du2024funcodec} is a unified neural audio codec typically operating at 16 kHz, designed to support a wide range of low bitrates (~1.5–12 kbps) via a modular RVQ-based framework. It emphasizes flexibility and generalization across compression settings and downstream speech applications.

\noindent\textbf{AudioDec.}
AudioDec~\cite{wu2023audiodec} proposes an end-to-end neural audio codec with a modular encoder–quantizer–decoder architecture and a two-stage training strategy that combines reconstruction and adversarial losses. It operates at a frame rate of 50 Hz and supports low-bitrate speech compression in 12.8 kbps, achieving low-latency, high-fidelity reconstruction suitable for streaming scenarios.

\noindent\textbf{AcadmiCodec.}
AcadmiCodec~\cite{yang2023hifi} introduces group residual vector quantization (GRVQ) to improve codebook utilization and reconstruction quality under constrained bitrates. It supports frame rates of 50 Hz and 75 Hz, and demonstrates strong performance across a wide bitrate range from 1.5 kbps to 24 kbps, while the accompanying AcademiCodec toolkit provides reproducible training pipelines and pretrained models for neural audio codecs.

\noindent\textbf{MP3, Opus, and AAC.}
To benchmark emotional trait preservation against established standards, we include widely used conventional codecs as baselines. MP3~\cite{brandenburg1994iso_mp3} provides a strong balance between compression efficiency and audio quality, Opus~\cite{valin2016high_opus} offers broad audio adaptability with low latency, and AAC~\cite{bosi1997iso_aac} is commonly adopted for high-fidelity streaming and broadcasting. These codecs serve as reference points for comparison with neural audio codecs.

\noindent\textbf{DAC.}
DAC~\cite{dac_kumar2023high} extends the VQGAN-style codec framework by projecting latent representations into a low-dimensional space prior to quantization, improving codebook utilization. We evaluate two reproduced variants: one employing three codebooks at a 25 Hz frame rate and another using a single codebook at 75 Hz. Both configurations operate at a 75 Hz token rate and achieve a bitrate of 0.75 kbps, providing a strong low-bitrate acoustic reconstruction baseline.

\noindent\textbf{SpeechTokenizer.}
SpeechTokenizer~\cite{xin2024speechtokenizer} enhances discrete speech representations by applying semantic distillation from HuBERT features to the first quantization layer. It operates at a 50 Hz token rate with two codebooks and is designed to improve linguistic modeling for downstream generation. We use the official released checkpoint in all experiments.

\noindent\textbf{Mimi.}
Mimi~\cite{mimi_defossez2024moshi} follows the hierarchical tokenization design of SpeechTokenizer but replaces the semantic teacher with WavLM representations. It employs eight codebooks, each of size 2,048, and operates at a low frame rate of 12.5 Hz, resulting in a bitrate of approximately 1.1 kbps. This configuration emphasizes compact tokenization with enhanced semantic supervision.

% \textbf{BiCodec.}
% BiCodec is a semantically enhanced tokenizer that uses a single-layer codebook to discretize speech based on wav2vec 2.0 features. It operates at a 50 Hz token rate with a codebook size of 8,192 and achieves a bitrate of 0.65 kbps. Its design prioritizes semantic abstraction while maintaining efficient compression.

\noindent\textbf{BigCodec.}
BigCodec~\cite{xin2024bigcodec} explores scaling single-codebook tokenization by increasing model capacity with sequential modules within convolutional architectures. It applies low-dimensional quantization to improve code utilization and operates at an 80 Hz token rate with a codebook size of 8,192, achieving a bitrate of 1.04 kbps.

\noindent\textbf{TAAE.} TAAE~\cite{parker2024scaling} is a transformer-based neural speech codec with nearly 1B parameters that adopts Finite Scalar Quantization (FSQ) instead of RVQ to enable ultra-low-bitrate compression.
Operating at 16 kHz with 25–50 Hz token rates, it achieves 400–700 bps while preserving high perceptual quality.

\noindent\textbf{L1asa.}
X-Codec 2~\cite{ye2025llasa} adopts a dual-encoder architecture consisting of a semantic encoder based on Wav2Vec2-BERT and an acoustic encoder for low-level features. The outputs of both encoders are concatenated before quantization. The tokenizer operates at a 50 Hz token rate with a large codebook of size 65,536, yielding a bitrate of 0.8 kbps. We use the official pretrained checkpoint.

\noindent\textbf{WavTokenizer.}
WavTokenizer~\cite{ji2024wavtokenizer} is a single-codebook tokenizer trained on approximately 800K hours of mixed-domain audio. It operates at a 75 Hz token rate with a codebook size of 4,096, resulting in a bitrate of 0.9 kbps. Its large-scale training enables robust performance across diverse acoustic conditions.

\subsection{Text-to-Speech Systems}

\noindent\textbf{F5-TTS.}
F5-TTS~\cite{chen2025f5} is a flow-matching-based text-to-speech system that directly maps text inputs to acoustic representations without explicit duration modeling. It serves as a strong non-autoregressive baseline for evaluating synthesis quality conditioned on discrete speech tokens.

\noindent\textbf{MaskGCT.}
MaskGCT~\cite{wang2024maskgct} is a large-scale masked generative TTS system that removes the need for explicit alignment between text and speech during training. It is trained on the Emilia dataset and relies on masked token prediction to generate speech, enabling flexible zero-shot synthesis.

\noindent\textbf{ARS.}
ARS~\cite{wang2024maskgct} is a cascaded autoregressive baseline combining an AR text-to-token model with a non-autoregressive codec-to-waveform decoder. It is also referred to as “AR + SoundStorm” and represents a common two-stage generation paradigm in codec-based TTS systems.

\noindent\textbf{CosyVoice 2.}
CosyVoice 2~\cite{du2024cosyvoice} is a large-scale zero-shot TTS system built upon an autoregressive language model initialized from Qwen2.5-0.5B-Instruct. It predicts speech tokens extracted by the CosyVoice 2 tokenizer, which operates at a 25 Hz token rate with a bitrate of approximately 0.325 kbps.

\noindent\textbf{FireRedTTS.}
FireRedTTS~\cite{guo2024fireredtts} is an autoregressive TTS system that predicts discrete speech codes extracted by the FireRedTTS tokenizer. The tokenizer employs HuBERT-based semantic features, a ResNet-style encoder, and a single codebook of size 16,384, with decoding performed via flow matching.

\noindent\textbf{SparkTTS.}
SparkTTS~\cite{wang2025spark} is an autoregressive TTS system initialized from Qwen2.5-0.5B-Instruct. It predicts speech tokens produced by the BiCodec tokenizer, enabling an evaluation of BiCodec representations in large-scale generative synthesis.

\noindent\textbf{Llasa.}
Llasa~\cite{ye2025llasa} is a large-scale TTS system built upon an autoregressive model initialized from Llama 3.2-1B. It predicts discrete speech tokens extracted by X-Codec 2, making it particularly relevant for assessing the compatibility of codec representations with large language models.

\section{Datasets}
\label{apx:datasets_detail}
\subsection{Reconstruction Task Datasets}
We train the codec on a multi-domain corpus of approximately 2.3K hours of audio,
designed to support high-fidelity reconstruction, multilingual robustness, and
emotion-aware representation learning. All recordings are resampled to 16 kHz.

\noindent\textbf{LibriSpeech}~\cite{panayotov2015librispeech} is used as the primary clean
English read-speech source. We include the \texttt{train-clean-100} and
\texttt{train-clean-360} splits, totaling approximately 460 hours. During
training, utterances are randomly cropped into 3-second segments. For evaluation,
we report reconstruction performance on \texttt{test-clean} and
\texttt{test-other}, representing clean and noisy conditions, respectively.

\noindent\textbf{VCTK}~\cite{VCTK_yamagishi2019cstr} provides multi-speaker English speech
with diverse accents. We use the full corpus (approximately 44 hours), originally
recorded at 48 kHz and downsampled to 16 kHz, to improve speaker and phonetic
diversity.

\noindent\textbf{AISHELL-3}~\cite{shi2020aishell} contributes Mandarin speech for
cross-lingual robustness. We use the full training set, comprising approximately
85 hours of high-quality recordings at 16 kHz.

\noindent\textbf{AudioSet}~\cite{gemmeke2017audio} introduces broad acoustic variability.
We incorporate a curated 1,000-hour subset covering diverse recording conditions
and background environments to improve generalization beyond clean studio speech.

\noindent\textbf{MSP-Podcast}~\cite{lotfian2017building_MSP-Podcast} is a large-scale
spontaneous conversational speech corpus totaling approximately 407 hours of
English audio from over 3,600 speakers. Collected from publicly available
podcasts, it covers diverse topics, speaking styles, and naturally occurring
emotional expressions, with multi-rater emotion annotations. All audio is
provided at 16~kHz, making it well suited for modeling real-world affective
speech.

\noindent\textbf{CMU-MOSEI}~\cite{CMU_MOSEI_bagher-zadeh-etal-2018-multimodal} is a
multimodal sentiment and emotion dataset derived from online videos. We use only
the speech modality, comprising approximately 65 hours of English audio with
utterance-level emotion annotations. This dataset enriches emotional
diversity under unconstrained, in-the-wild recording conditions.

\noindent\textbf{EmoVoiceDB}~\cite{yang2025emovoice} is a high-quality English emotional speech dataset designed for fine-grained emotion modeling and evaluation. It contains 40 hours of emotionally expressive speech, comprising over 22,000 utterances annotated with natural language emotion descriptions and covering seven core emotion categories. The dataset includes diverse speaker timbres and expressive styles. All audio is provided at a 16~kHz sampling rate. In our work, EmoVoiceDB is used exclusively as an evaluation benchmark to assess emotional consistency and preservation on reconstructed speech.

\subsection{Benchmark for SER Evaluation}
\textbf{EMO-SUPERB}~\cite{wu2024_emo_superb} aggregates six public SER datasets, including IMPROV, CREMA-D, MSP-Podcast, BIIC-Podcast, IEMOCAP, and NNIME, covering acted, improvised, and real-world emotional speech. In total, the benchmark contains approximately 414 hours of audio from over 2,300 speakers across English and Mandarin. All audio is resampled to 16 kHz and partitioned using speaker-independent splits to ensure robust evaluation.
We follow the official EMO-SUPERB data splits and evaluate emotion recognition performance on speech reconstructed by each codec.

\subsection{TTS Task Datasets}

\noindent\textbf{LibriTTS}~\cite{zen2019libritts} is used to train the downstream text-to-speech
token prediction models. LibriTTS is an English read-speech corpus derived from
LibriSpeech, providing paired text and speech data with high recording quality.
We use the official \texttt{train} and \texttt{dev} splits, totaling approximately
585 hours of audio. All recordings are resampled to 16 kHz. This dataset serves
as the primary source for learning text-conditioned generation over codec tokens.

\noindent\textbf{SECAP}~\cite{xu2024secap} is a speech emotion captioning dataset containing emotionally expressive utterances annotated with natural language emotion descriptions.
The dataset consists of approximately 40 hours of speech from 7 speakers, originally recorded at 24 kHz.
We resample the audio to 16 kHz and use it solely for evaluation, focusing on emotional expressiveness and semantic alignment in generated speech rather than caption generation.

\section{Subjective Reconstruction Evaluation}
\label{apx:subjective_reconstruction}
\begin{figure*}[t]
    \centering
    \begin{minipage}[t]{0.32\textwidth}
        \centering
        \includegraphics[width=\linewidth]{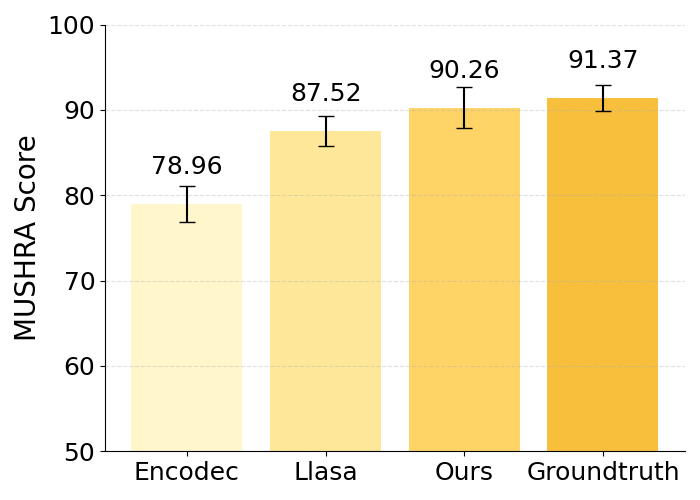}
        %\caption{Figure (a): Subjective MUSHRA evaluation comparing Encodec, Llasa, our model, and ground-truth recordings. Our model achieves the highest perceptual quality among neural codecs, approaching the ground-truth reference.}
    \end{minipage}
    \hfill
    \begin{minipage}[t]{0.32\textwidth}
        \centering
        \includegraphics[width=\linewidth]{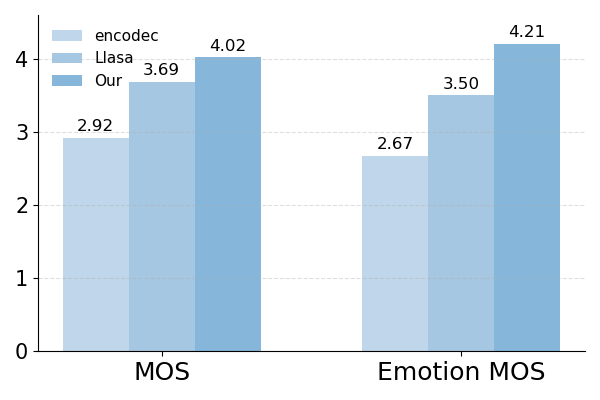}
        %\caption{Figure (b): Mean Opinion Score (MOS) and Emotion-MOS results across models. Our method improves both perceived audio quality and emotional expressiveness compared to Encodec and Llasa.}
    \end{minipage}
    \hfill
    \begin{minipage}[t]{0.33\textwidth}
        \centering
        %\vspace{-1pt}
        \includegraphics[width=\linewidth]{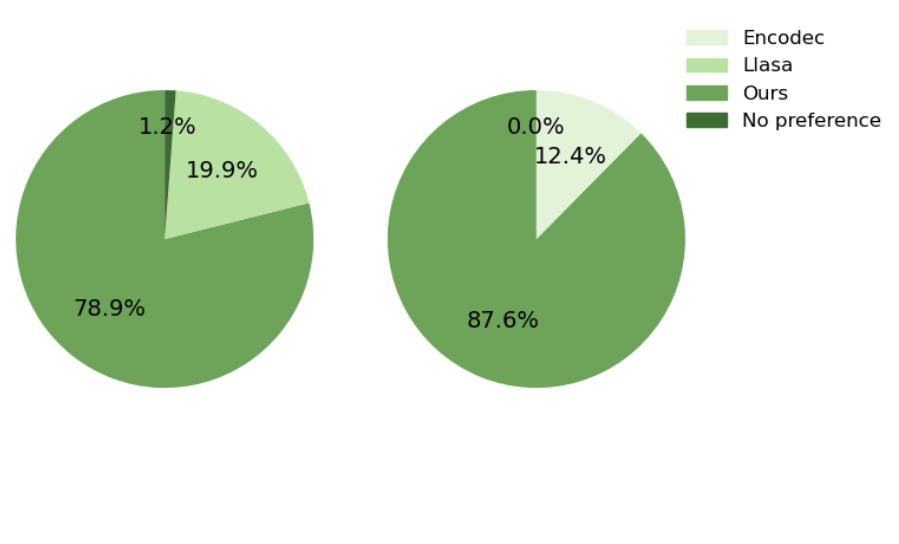}
        %\caption{Figure (c): AB preference test showing listener choices between our model and competing systems. The majority of participants consistently prefer our synthesized speech over Encodec and Llasa.}
    \end{minipage}
    \caption{\textbf{Reconstruction subjective evaluation results} across three complementary settings.
(a) MUSHRA scores comparing Encodec, Llasa, our method, and ground-truth recordings, evaluating overall perceptual quality under a reference-based protocol.
(b) MOS and Emotion-MOS results assessing naturalness and affective expressiveness across competing systems.
(c) AB-preference results measuring pairwise perceptual preference and emotional preference.}
    \label{fig:subjective_all}
\end{figure*}

\paragraph{MUSHRA Evaluation.}
We conduct a MUSHRA evaluation to assess overall perceptual quality under a
reference-based listening protocol~\cite{encodec_defossez2022high}. Ground-truth recordings are provided as
upper anchors, with EnCodec and Llasa included as competitive baselines.
A total of 24 listeners participated in the study, and each listener
evaluated randomly selected utterances on a 0--100 scale following standard
MUSHRA guidelines.
As shown in Figure~\ref{fig:subjective_all} (a), our method achieves a mean MUSHRA score of 90.26,
substantially outperforming EnCodec (78.96) and Llasa (87.52), and closely
approaching the ground-truth upper bound (91.37). These results indicate that
the proposed emotion-guided codec significantly improves perceptual
reconstruction quality.

\paragraph{MOS and Emotion-MOS Evaluation.}
We further assess speech naturalness and affective expressiveness using Mean
Opinion Score (MOS) and Emotion-MOS evaluations by fowllowing~\cite{gao2025emo}. Listeners rated each utterance
on a 1--5 scale for overall naturalness and emotional expressiveness,
respectively. Each sample was evaluated by 24 independent raters, and
scores were averaged across listeners.
As illustrated in Figure~\ref{fig:subjective_all} (b), our method achieves the highest MOS (4.02) and
Emotion-MOS (4.21), outperforming both EnCodec (2.92 / 2.67) and Llasa
(3.69 / 3.50). The larger improvement in Emotion-MOS highlights the benefit of
explicit emotion modeling in discrete representation learning.

\paragraph{AB Preference Evaluation.}
We additionally conduct AB preference tests to directly compare perceptual and
emotional preference between systems. In each trial, listeners were presented
with paired samples and asked to indicate overall quality preference and
emotional preference. Each comparison was judged by 24 listeners, with
randomized order and balanced pairings.
As shown in Figure~\ref{fig:tts_subjective_all} (c), our method is preferred over EnCodec in 78.9\% of overall
quality judgments and 87.6\% of emotional preference judgments, demonstrating a
strong listener preference for the proposed emotion-aware codec.

\section{Subjective Evaluation for TTS}
\label{apx:subjective_tts}
\begin{figure*}[t]
    \centering
    \begin{minipage}[t]{0.38\textwidth}
        \centering
        \includegraphics[width=\linewidth]{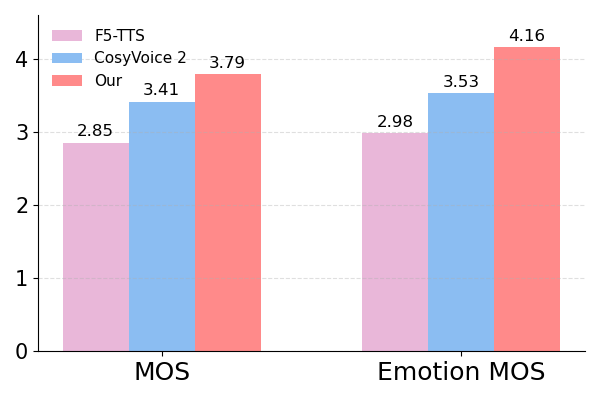}
        %\caption{Figure (b): Mean Opinion Score (MOS) and Emotion-MOS results across models. Our method improves both perceived audio quality and emotional expressiveness compared to Encodec and Llasa.}
    \end{minipage}
    \hfill
    \begin{minipage}[t]{0.52\textwidth}
        \centering
        %\vspace{-1pt}
        \includegraphics[width=\linewidth]{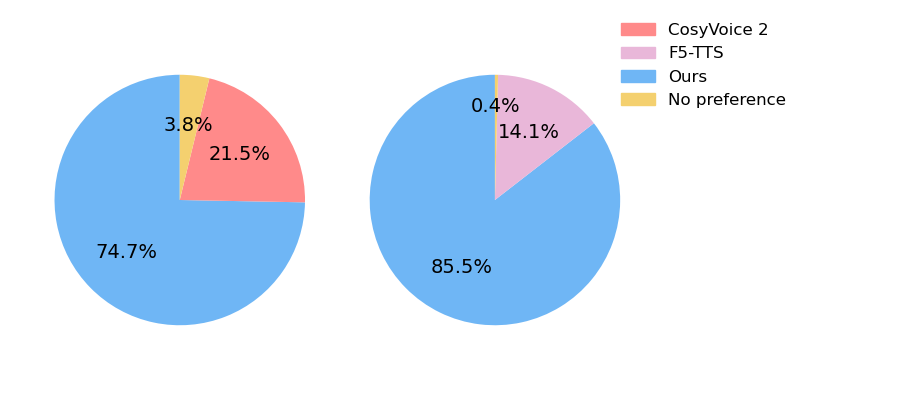}
        %\caption{Figure (c): AB preference test showing listener choices between our model and competing systems. The majority of participants consistently prefer our synthesized speech over Encodec and Llasa.}
    \end{minipage}
    \caption{\textbf{Text To Speech subjective evaluation results} across two complementary settings.
(a) MOS and Emotion-MOS results assessing naturalness and affective expressiveness across competing systems.
(b) AB-preference results measuring pairwise perceptual preference and emotional preference.}
    \label{fig:tts_subjective_all}
\end{figure*}

We conduct subjective listening tests to evaluate perceptual quality and emotional expressiveness of the generated speech using three complementary protocols: Mean Opinion Score (MOS), Emotion Mean Opinion Score (Emotion MOS), and AB preference tests. A total of 24 listeners participated in all evaluations. MOS assesses audio quality and naturalness on a five-point Likert scale ranging from 1 (bad) to 5 (excellent), while Emotion MOS measures the perceived similarity between the emotion expressed in the generated speech and that of the reference audio, rated from 1 (not at all similar) to 5 (extremely similar). For MOS and Emotion MOS evaluations, listeners were asked to rate 30 utterances per system, with emotion categories evenly balanced across samples. The evaluated systems include F5-TTS, CosyVoice~2, and our method. As shown in Figure~\ref{fig:tts_subjective_all} (a), our method achieves the highest scores on both MOS and Emotion MOS. Specifically, our system attains a MOS of 3.79, outperforming CosyVoice~2 (3.41) and F5-TTS (2.85). A similar trend is observed for Emotion MOS, where our method reaches 4.16, compared to 3.53 for CosyVoice~2 and 2.98 for F5-TTS. The larger margin on Emotion MOS indicates that our approach yields more faithful emotional expression, while also improving overall perceptual quality. We further conduct AB preference tests to directly compare listener preferences between systems. In each trial, listeners were presented with paired samples generated by two systems and asked to select the preferred one based on overall quality and emotional expressiveness, or indicate no preference. Two pairwise comparisons are performed: CosyVoice~2 versus our method and F5-TTS versus our method, using emotion-balanced samples. As shown in Figure~\ref{fig:tts_subjective_all} (b), listeners prefer our method in 74.7\% of comparisons against CosyVoice~2 and 85.5\% against F5-TTS, while the proportion of neutral responses remains below 4\% in both cases. These results demonstrate a strong and consistent subjective preference for our method, confirming its advantages in both perceptual naturalness and emotional expressiveness during speech synthesis.

\section{Fine-Grained Ablation Analysis}
\label{apx:fine-grained ablation}
\subsection{Emotion--Semantic Guided Latent}
\begin{table*}[ht]
\centering
\caption{
\textbf{Ablation Study of Emotion--Semantic Guided Latent Modulation}.
Cross variants incorporate cross-modal attention between emotion and semantic signals,
while Self variants apply self-attention.
Before applies attention prior to projection into the encoder latent space,
whereas After applies attention post-projection.
None denotes direct projection without attention. Results are averaged over three random seeds.
}
\label{tab:evaluation_ablation}
\small
\resizebox{\textwidth}{!}{
\begin{tabular}{l ccc ccc ccc}
\toprule
\multirow{2}{*}{Attn--Proj Type}
& \multicolumn{3}{c}{Emotion Consistency}
& \multicolumn{3}{c}{Content Preservation}
& \multicolumn{3}{c}{Speech Naturalness} \\
\cmidrule(lr){2-4} \cmidrule(lr){5-7} \cmidrule(lr){8-10}
& Emo SIM$\uparrow$ & Pros SIM$\uparrow$ & Recall$\uparrow$
& WER$\downarrow$ & WIL$\downarrow$ & LSD$\downarrow$
& MSEP$\downarrow$ & PESQ$\uparrow$ & UTMOS$\uparrow$ \\
\midrule

Ours (Cross-Attn-Before)
& \textbf{0.94} & \textbf{0.86} & \textbf{0.48}
& \underline{4.15} &  \underline{6.43} & \textbf{0.78}
& \textbf{19.21} & \textbf{3.04} & \textbf{3.68} \\

None-Attn
& 0.90 & 0.79 & 0.35
& 4.58 & 6.60 & 1.05
& 34.85 & 2.67 &  \underline{3.52} \\
\midrule

\quad w/ Sem-only
& 0.89 & 0.79 & 0.34
& 4.67 & 6.61 & 0.83
& 25.91 & 2.75 & 3.48 \\

\quad w/ Emo-only
& 0.91 & 0.81 & 0.41
& 4.72 & 6.65 & 0.82
& 26.40 & 2.69 & 3.39 \\
\midrule

\quad w/ Self-Attn-After
& 0.90 & 0.82 & 0.40
& 4.67 & 6.61 & 0.84
& 25.10 & 2.71 & 3.34 \\

\quad w/ Self-Attn-Before
& \underline{0.92} & \underline{0.84} & \underline{0.44}
& \textbf{4.09} & \textbf{6.36} & \underline{0.80}
& \underline{22.85} &  \underline{2.89} & 3.51 \\

\quad w/ Cross-Attn-After
& 0.91 & 0.83 & 0.42
& 4.17 & 6.70 & 0.89
& 23.90 & 2.82 & 3.46 \\

\bottomrule
\end{tabular}
}
\end{table*}

Table~\ref{tab:evaluation_ablation} reports a fine-grained ablation study on
attention--projection strategies for emotion--semantic guided latent modulation,
evaluated along emotion consistency, content preservation, and speech naturalness.

The None--Attn variant removes attention and directly injects auxiliary signals by
setting $\mathbf{h}^{\mathrm{emo}}_t=\tilde{\mathbf{e}}_t$ and
$\mathbf{h}^{\mathrm{sem}}_t=\tilde{\mathbf{s}}_t$, yielding a unified latent
$\mathbf{z}^{\mathrm{uni}}_t=\mathbf{z}_t+W_m\tilde{\mathbf{e}}_t+W_m\tilde{\mathbf{s}}_t$.
This naive formulation preserves coarse linguistic content with WER 4.58 and WIL
6.60, but exhibits clear degradation in emotional modeling, as reflected by low
Emo SIM 0.90 and recall 0.35. In addition, severe spectral distortion is observed,
with LSD increasing to 1.05 and MSEP reaching 34.85, indicating that direct feature
injection fails to maintain emotion-related structure under quantization.

Single-source supervision partially alleviates this limitation.
In the Sem-only variant, emotion modulation is disabled by setting
$\mathbf{h}^{\mathrm{emo}}_t=\mathbf{0}$ while retaining
$\mathbf{h}^{\mathrm{sem}}_t=\tilde{\mathbf{s}}_t$.
This configuration improves spectral stability compared to None--Attn, reducing
LSD to 0.83 and MSEP to 25.91, but yields limited gains in emotion consistency,
with Emo SIM remaining at 0.89 and recall at 0.34.
Conversely, the Emo-only variant sets
$\mathbf{h}^{\mathrm{emo}}_t=\tilde{\mathbf{e}}_t$ and
$\mathbf{h}^{\mathrm{sem}}_t=\mathbf{0}$, leading to stronger emotion preservation
with recall improving to 0.41 and Emo SIM to 0.91, but at the cost of weaker content
robustness and higher spectral distortion.
These results confirm that emotion and semantic cues provide complementary but
insufficient supervision when used independently.

Introducing attention mechanisms further improves representation quality.
Self-attention variants replace direct projections with
$\mathbf{h}^{\mathrm{emo}}_t=\mathrm{SelfAttn}(\tilde{\mathbf{E}})_t$ and
$\mathbf{h}^{\mathrm{sem}}_t=\mathrm{SelfAttn}(\tilde{\mathbf{S}})_t$.
Among them, Self--Attn--Before, which applies attention prior to projection,
achieves stronger emotion consistency with recall 0.44 and improved content
preservation with WER 4.09 and WIL 6.36 compared to Self--Attn--After.
It also reduces spectral distortion, attaining LSD 0.80 and MSEP 22.85,
suggesting that early interaction in the original feature space is more effective
than post-projection refinement. Cross-modal attention yields the most consistent gains by explicitly conditioning
acoustic latents on auxiliary signals.
The Cross--Attn--After variant applies cross attention after projection and provides
moderate improvements in emotion consistency and perceptual quality.
In contrast, the full Cross--Attn--Before configuration applies cross attention
prior to projection, enabling richer interaction between acoustic, emotional, and
semantic representations in a shared latent space.
This design achieves the best overall balance, attaining the highest Emo SIM 0.94
and Pros SIM 0.86, while substantially reducing spectral distortion with LSD 0.78
and MSEP 19.21.
Corresponding gains in PESQ 3.04 and UTMOS 3.68 further demonstrate that emotion--
semantic guided cross-modal interaction before projection is critical for
preserving emotional expressiveness without compromising content fidelity or
perceptual naturalness.
\begin{table*}[ht]
\centering
\caption{\textbf{Ablation Study of Relation-Preserving Emotional–Semantic Distillation}.The full RP-Distill is compared with several ablated variants, including None-Distill without relational supervision, Sem-only using semantic relational distillation only, Emo-only using emotion relational distillation only, and Feature Distill applying feature-level distillation without relational constraints. Results are averaged over three random seeds.
}
\label{tab:evaluation_ablation_distill}
\small
\resizebox{\textwidth}{!}{
\begin{tabular}{l ccc ccc ccc}
\toprule
\multirow{2}{*}{Distillation Type}
& \multicolumn{3}{c}{Emotion Consistency}
& \multicolumn{3}{c}{Content Preservation}
& \multicolumn{3}{c}{Speech Naturalness} \\
\cmidrule(lr){2-4} \cmidrule(lr){5-7} \cmidrule(lr){8-10}
& Emo SIM$\uparrow$ & Pros SIM$\uparrow$ & Recall$\uparrow$
& WER$\downarrow$ & WIL$\downarrow$ & LSD$\downarrow$
& MSEP$\downarrow$ & PESQ$\uparrow$ & UTMOS$\uparrow$ \\
\midrule

Ours (RP-Distill)
& \textbf{0.94} & \textbf{0.86} & \textbf{0.48}
& \textbf{4.15} & \textbf{6.43} & \textbf{0.78}
& \textbf{19.21} & \textbf{3.04} & \textbf{3.68} \\

None-Distill
& 0.92 & 0.81 & 0.40
& 4.62 & 6.78 & 1.09
& 40.16 & 2.49 & 3.44 \\
\midrule

\quad w/ Sem-only
& 0.92 & 0.82 & 0.42
& \underline{4.34} & \underline{6.53} & \underline{0.81}
& 22.58 & \underline{2.83} & \underline{3.52} \\

\quad w/ Emo-only
& \underline{0.93} & \underline{0.84} & \underline{0.45}
& 4.60 & 6.65 & 0.83
& 25.14 & 2.66 & 3.46 \\
\midrule

\quad w/ Feature Distill
& 0.92 & 0.82 & 0.42
& 4.42 & 6.59 & 0.82
& \underline{22.47} & 2.80 & 3.49 \\

\bottomrule
\end{tabular}
}
\end{table*}
\subsection{Relation-Preserving Emotional--Semantic Distillation}

Table~\ref{tab:evaluation_ablation_distill} presents a fine-grained ablation study of the proposed
relation-preserving emotional--semantic distillation, evaluated from emotion consistency,
content preservation, and speech naturalness.

The \textit{None-Distill} variant removes relational supervision entirely, allowing unified
latents to be optimized solely by reconstruction objectives. As a result, performance degrades
substantially across all dimensions, with notable drops in emotion consistency
(Emo SIM 0.92, Recall 0.40) and severe spectral distortion
(LSD 1.09, MSEP 40.16). This confirms that residual vector quantization alone is insufficient
to preserve relational structure across emotional and semantic spaces.

The \textit{Sem-only} variant introduces relational distillation using only semantic teacher
relations by disabling emotional supervision. The corresponding objective is defined as
\begin{equation}
\mathcal{L}_{\mathrm{rela}}^{\mathrm{sem}}
=
\frac{1}{T'^2}
\sum_{t=1}^{T'}
\sum_{t'=1}^{T'}
\Big[
\beta \, d\!\left(r^{\mathrm{uni}}_{t,t'}, r^{\mathrm{sem}}_{t,t'}\right)
\Big],
\quad \alpha = 0.
\tag{6}
\end{equation}
This setting yields clear improvements over \textit{None-Distill} in content preservation
(WER 4.34 vs.\ 4.62, LSD 0.81 vs.\ 1.09) and speech naturalness
(PESQ 2.83, UTMOS 3.52). However, gains in emotion consistency remain limited
(recall 0.42), indicating that semantic relations alone cannot fully recover affective structure.

The \textit{Emo-only} variant instead distills relational structure exclusively from the emotion
teacher by disabling semantic supervision:
\begin{equation}
\mathcal{L}_{\mathrm{rela}}^{\mathrm{emo}}
=
\frac{1}{T'^2}
\sum_{t=1}^{T'}
\sum_{t'=1}^{T'}
\Big[
\alpha \, d\!\left(r^{\mathrm{uni}}_{t,t'}, r^{\mathrm{emo}}_{t,t'}\right)
\Big],
\quad \beta = 0.
\tag{7}
\end{equation}
This configuration substantially improves emotion consistency
(Emo SIM 0.93, recall 0.45), but provides weaker benefits for content fidelity and spectral
stability, reflected by higher WER (4.60) and LSD (0.83). These results suggest that emotional
relations alone are insufficient to stabilize fine-grained acoustic structure under quantization.

The \textit{Feature Distill} variant replaces relational constraints with direct feature matching
between unified latents and teacher representations:
\begin{equation}
\mathcal{L}_{\mathrm{feat}}
=
\frac{1}{T'}
\sum_{t=1}^{T'}
\left(
\left\lVert \mathbf{z}^{\mathrm{uni}}_t - \mathbf{e}_t \right\rVert_2^2
+
\left\lVert \mathbf{z}^{\mathrm{uni}}_t - \mathbf{s}_t \right\rVert_2^2
\right).
\tag{8}
\end{equation}
While feature-level distillation directly aligns unified latents with projected teacher embeddings and improves overall stability compared to \textit{None-Distill},
its performance remains inferior to relation-preserving supervision, particularly in emotion
consistency (recall 0.42) and perceptual quality (PESQ 2.80).
In contrast, the full relation-preserving emotional--semantic distillation achieves the best
overall balance, simultaneously maximizing emotion consistency, reducing spectral distortion,
and improving perceptual naturalness. These results demonstrate that preserving pairwise
relational geometry across both emotion and semantic spaces is critical for maintaining expressive
and content-faithful discrete representations under quantization.
\subsection{Emotion-Weighted Semantic Alignment}
\begin{table*}[t]
\centering
\caption{\textbf{Ablation Study of Emotion-Weighted Semantic Alignment.} The full model employs emotion-aware frame weighting during semantic alignment.
Sem-only Align removes emotion guidance and relies solely on semantic supervision.
Uniform-Scaled Weighted applies uniform weighting without considering emotional variation. Results are averaged over three random seeds.}
\label{tab:evaluation_ablation_align}
\small
\resizebox{\textwidth}{!}{
\begin{tabular}{l ccc ccc ccc}
\toprule
\multirow{2}{*}{Distillation Type}
& \multicolumn{3}{c}{Emotion Consistency}
& \multicolumn{3}{c}{Content Preservation}
& \multicolumn{3}{c}{Speech Naturalness} \\
\cmidrule(lr){2-4} \cmidrule(lr){5-7} \cmidrule(lr){8-10}
& Emo SIM$\uparrow$ & Pros SIM$\uparrow$ & Recall$\uparrow$
& WER$\downarrow$ & WIL$\downarrow$ & LSD$\downarrow$
& MSEP$\downarrow$ & PESQ$\uparrow$ & UTMOS$\uparrow$ \\
\midrule

Ours (EW-Align)
& \textbf{0.94} & \textbf{0.86} & \textbf{0.48}
& \textbf{4.15} & \textbf{6.43} & \textbf{0.78}
& \textbf{19.21} & \textbf{3.04} & \textbf{3.68} \\

None-Align
& 0.93 & 0.82 & 0.43
& 4.51 & 6.67 & 1.07
& 36.59 & 2.58 & 3.49 \\
\midrule

\quad w/ Sem-only Align
& \underline{0.93} & \underline{0.83} & \underline{0.44}
& \underline{4.32} & \underline{6.49} & \underline{0.83}
& \underline{24.78} & \underline{2.84} & \underline{3.56} \\

\quad w/ Uniform-Scaled Weighted
& 0.92 & 0.82 & 0.42
& 4.48 & 6.61 & 0.97
& 31.35 & 2.66 & 3.50 \\

\bottomrule
\end{tabular}
}
\end{table*}
\begin{table*}[t]
\centering
\caption{\textbf{Ablation of RVQ layer selection.}
First Layer supervises only the first RVQ layer $\mathbf{Q}^{(1)}$, whereas Early
Layers and All Layers progressively include deeper quantization stages.
Across all evaluation dimensions, performance degrades as supervision extends
to deeper RVQ layers, indicating that the earliest RVQ layer captures the most
compact and informative representation for emotion-aware speech modeling.}
\label{tab:evaluation_ablation_layer}
\small
\resizebox{\textwidth}{!}{
\begin{tabular}{l ccc ccc ccc}
\toprule
\multirow{2}{*}{RVQ Layer Selection}
& \multicolumn{3}{c}{Emotion Consistency}
& \multicolumn{3}{c}{Content Preservation}
& \multicolumn{3}{c}{Speech Naturalness} \\
\cmidrule(lr){2-4} \cmidrule(lr){5-7} \cmidrule(lr){8-10}
& Emo SIM$\uparrow$ & Pros SIM$\uparrow$ & Recall$\uparrow$
& WER$\downarrow$ & WIL$\downarrow$ & LSD$\downarrow$
& MSEP$\downarrow$ & PESQ$\uparrow$ & UTMOS$\uparrow$ \\
\midrule

\textbf{Ours}: First Layer ($\mathbf{Q}^{(1)}$)
& \textbf{0.94} & \textbf{0.86} & \textbf{0.48}
& \textbf{4.15} & \textbf{6.43} & \textbf{0.78}
& \textbf{19.21} & \textbf{3.04} & \textbf{3.68} \\

\midrule
\quad w/ Early Layers ($\mathbf{Q}^{(1:4)}$)
& \underline{0.93} & \underline{0.84} & \underline{0.45}
& \underline{4.22} & \underline{6.42} & \underline{0.83}
& \underline{22.85} & \underline{2.89} & \underline{3.61} \\

\quad w/ All Layers ($\mathbf{Q}^{(1:8)}$)
& 0.91 & 0.82 & 0.41
& 4.38 & 6.65 & 0.95
& 26.90 & 2.63 & 3.52 \\

\bottomrule
\end{tabular}
}
\end{table*}
In Table~\ref{tab:evaluation_ablation_align}, this ablation study examines how emotion-aware weighting affects semantic alignment under discrete quantization.
By progressively removing or simplifying the emotion-dependent weighting mechanism, the three variants isolate the contributions of explicit emotion guidance, semantic-only alignment, and uniform scaling to representation robustness and expressiveness.
The \textit{None-Align} variant completely removes the emotion-aware semantic alignment objective, aiming to evaluate codec behavior when discrete representations are trained without explicit semantic alignment.
In this setting, the alignment loss $\mathcal{L}_{\mathrm{align}}$ is disabled, and optimization relies solely on reconstruction and prior supervisory objectives.
As reported in Table~10, removing alignment leads to noticeable degradation in both spectral stability and perceptual quality, with LSD increasing to 1.07 and MSEP rising sharply to 36.59.
Although emotion consistency remains moderately high in terms of Emo SIM (0.93), recall drops to 0.43, indicating weaker preservation of emotion-related temporal structure.
These results suggest that semantic alignment plays a crucial role in stabilizing discrete representations under quantization, particularly for emotionally expressive speech.

The \textit{Sem-only Align } variant introduces semantic alignment while removing emotion-aware weighting, in order to isolate the effect of semantic supervision alone.
The alignment objective reduces to a uniformly weighted formulation:
\begin{equation}
\mathcal{L}_{\mathrm{align}}^{\mathrm{Sem}}
=
-\frac{1}{T'}
\sum_{t=1}^{T'}
\log \sigma\!\left(\cos(Q_t^{(1)}, C_t^*)\right),
\tag{9}
\end{equation}
where all frames contribute equally regardless of emotional variation.
Compared to None-Align, Sem-only Align substantially improves content preservation and spectral fidelity, reducing WER to 4.32, WIL to 6.49, and LSD to 0.83, with a large decrease in MSEP from 36.59 to 24.78.
However, gains in emotion consistency remain limited, with recall reaching 0.44 and Pros SIM 0.83.
This indicates that semantic alignment improves intelligibility and acoustic stability, but lacks the ability to selectively protect emotionally salient regions.

The \textit{Uniform-Scaled Weighted} variant retains emotion-derived weighting but removes temporal adaptivity by replacing frame-wise weights with a global scalar.
Specifically,the frame-level emotion weights $\{\gamma_t\}_{t=1}^{T'}$ are averaged into a single constant scaling factor
$\bar{\gamma} = \frac{1}{T'} \sum_{t=1}^{T'} \gamma_t$,
which is then applied uniformly across all frames, removing temporal emotion variation while preserving the overall weighting magnitude.
\begin{equation}
\mathcal{L}_{\mathrm{align}}^{\mathrm{Uni}}
=
-\frac{1}{T'}
\sum_{t=1}^{T'}
\bar{\gamma}
\log \sigma\!\left(\cos(Q_t^{(1)}, C_t^*)\right).
\tag{10}
\end{equation}
This design preserves overall emotion awareness while discarding frame-level discrimination.
As shown in Table~10, Uniform-Scaled Weighted improves upon None-Align in content preservation, reducing WER to 4.48 and MSEP to 31.35, and yields slightly better perceptual quality with UTMOS 3.50.
However, emotion consistency degrades compared to Sem-only Align, with recall decreasing to 0.42 and LSD remaining relatively high at 0.97.
These results demonstrate that emotion-aware weighting alone is insufficient without temporal adaptivity, highlighting the importance of frame-level emotion-sensitive alignment.
\subsection{Effect of RVQ layer selection }
Table~\ref{tab:evaluation_ablation_layer} examines the impact of supervising different depths of RVQ layers.
In our framework, both the relation-preserving distillation stage
(\S\ref{stage:2}) and the emotion-weighted semantic alignment stage
(\S\ref{stage:3}) apply supervision exclusively to the first RVQ layer
$\mathbf{Q}^{(1)}=\{\mathbf{Q}^{(1)}_t\}_{t=1}^{T'} \in \mathbb{R}^{T' \times D}$,
which consistently achieves the best performance across emotion consistency,
content preservation, and speech naturalness metrics.
To study the effect of incorporating deeper quantization stages, we further
consider early-layer supervision
$\mathbf{Q}^{(1:4)} = \frac{1}{4}\sum_{i=1}^{4}\mathbf{Q}^{(i)} \in \mathbb{R}^{T' \times D}$
and full-layer supervision
$\mathbf{Q}^{(1:8)} = \frac{1}{8}\sum_{i=1}^{8}\mathbf{Q}^{(i)} \in \mathbb{R}^{T' \times D}$,
where representations from multiple RVQ layers are averaged prior to supervision.
Aggregating early RVQ layers leads to moderate degradation, while extending
supervision to all RVQ layers further degrades performance across all evaluation
dimensions. This monotonic trend suggests that the first RVQ layer captures the most compact
and informative structure for emotion-aware speech modeling, whereas deeper RVQ
layers predominantly encode residual acoustic details that are less stable and
more sensitive to quantization noise.
These observations align with prior findings in SpeechTokenizer, which show that
RVQ-1 exhibits stronger structural alignment and higher information efficiency
than deeper quantization layers~\cite{xin2024speechtokenizer}.
Overall, the results indicate that focusing supervision on the earliest RVQ layer
is crucial for preserving emotionally and semantically salient structure in
discrete speech representations.
\subsection{Emotion Encoder Selection }

To assess the effect of emotion encoder choice and the robustness of our framework, we evaluate representative encoders that balance emotion modeling and generalization. As shown in Table~12, CLAP-LAION~\cite{clap_wu2023large} achieves the most balanced overall performance across all metrics, benefiting from its large-scale, language-grounded contrastive training, which promotes both affective generalization and semantic consistency. It attains strong emotion preservation (Emo SIM = 0.94, Pros SIM = 0.86, recall = 0.48) while maintaining content fidelity (WER = 4.15, WIL = 6.43) and speech naturalness (PESQ = 3.04, UTMOS = 3.68). In contrast, CLEP-DG~\cite{shi2025clep} yields a slightly higher Emo SIM (0.95) but shows degraded prosodic consistency and increased distortion, reflecting its reliance on emotion-specific training that limits semantic generalization. CLAP-MST~\cite{elizalde2023clap}, despite sharing a similar training paradigm, underperforms CLAP-LAION due to its smaller training scale. AudioCLIP~\cite{guzhov2022audioclip} consistently lags behind, particularly in emotion consistency and naturalness, indicating limited suitability for emotion-aware codec supervision. Overall, these results highlight the advantage of large-scale, language-aligned emotion encoders for balanced emotion, content, and prosody preservation. 

\label{apx:emotion_encoder}
\begin{table*}[t]
\centering
\caption{\textbf{Ablation of Emotion Encoder Selection.}
We evaluate representative emotion encoders designed to balance
generalization and emotion preservation. CLAP-LAION demonstrates superior
emotion generalization while maintaining strong semantic consistency and
prosodic fluency. Results are averaged over three random seeds.}
\label{tab:evaluation_ablation_emotion_encoder}
\small
\resizebox{\textwidth}{!}{
\begin{tabular}{l ccc ccc ccc}
\toprule
\multirow{2}{*}{Emo-Encoder Selection}
& \multicolumn{3}{c}{Emotion Consistency}
& \multicolumn{3}{c}{Content Preservation}
& \multicolumn{3}{c}{Speech Naturalness} \\
\cmidrule(lr){2-4} \cmidrule(lr){5-7} \cmidrule(lr){8-10}
& Emo SIM$\uparrow$ & Pros SIM$\uparrow$ & Recall$\uparrow$
& WER$\downarrow$ & WIL$\downarrow$ & LSD$\downarrow$
& MSEP$\downarrow$ & PESQ$\uparrow$ & UTMOS$\uparrow$ \\
\midrule

 Ours (CLAP-LAION)
& \underline{0.94} & \textbf{0.86} & \textbf{0.48}
& \textbf{4.15} & \textbf{6.43} & \textbf{0.78}
& \textbf{19.21} & \textbf{3.04} & \textbf{3.68} \\
\midrule
\quad w/ CLEP-DG
& \textbf{0.95} & \underline{0.84} & \underline{0.47}
& 4.53 & 6.93 & 0.90
& 21.12 & 2.86 & 3.57 \\

\quad w/ CLAP-MST
& 0.92 & 0.84 & 0.44
& \underline{4.26} & \underline{6.55} & \underline{0.85}
& \underline{20.42} & \underline{2.97} & \underline{3.61} \\

\quad w/ AudioClip
& 0.90 & 0.82 & 0.40
& 4.82 & 7.31 & 0.94
& 23.03 & 2.78 & 3.49 \\

\bottomrule
\end{tabular}
}
\end{table*}

\section{Model Details}
\label{apx:model_details}
\subsection{Encoder and Decoder Architecture}
The proposed codec adopts a standard neural audio codec backbone widely used in
recent work on discrete speech representations and neural tokenizers
\cite{encodec_defossez2022high,zeghidour2021soundstream,xin2024speechtokenizer}.
The encoder processes raw waveforms through a hierarchical convolutional
architecture to produce temporally downsampled latent representations~\cite{xiao2024ucil,peng2024dark}. Specifically, the encoder begins with a one-dimensional convolutional layer with
32 channels and a kernel size of 7, followed by four stacked residual convolutional
blocks. Each block contains two convolutional layers with kernel size $(3,1)$ and
unit dilation, a residual connection, and a strided convolution for temporal
downsampling. The stride factors across the four blocks are set to $(2,4,5,8)$,
with kernel sizes twice the corresponding stride. The number of channels is doubled
at each downsampling stage to progressively increase representational capacity. To capture long-range temporal dependencies, the convolutional backbone is followed
by a two-layer bidirectional LSTM. A final one-dimensional convolution with kernel
size 7 projects the hidden states to the target embedding dimension $D$. ELU~\cite{clevert2015fast}
activations are used throughout the network, and either layer
normalization or weight normalization
is applied depending on the layer type. The decoder mirrors the encoder architecture in reverse order. Strided convolutions
are replaced with transposed convolutions to restore temporal resolution, and LSTM
layers are used to reconstruct long-range structure. The decoder outputs the
reconstructed waveform at the original sampling rate.

\subsection{Residual Vector Quantization}

The continuous encoder outputs are discretized using a Residual Vector Quantization
(RVQ) module following prior neural codec designs
\cite{encodec_defossez2022high,xin2024speechtokenizer}. RVQ consists of a sequence of
$K$ codebooks, where each codebook quantizes the residual error left by the previous
ones, enabling progressive refinement of the discrete representation. In our implementation, the encoder latent tensor of shape $[B, D, T]$ is quantized
using $K=8$ residual codebooks, each containing 1024 entries. At each quantization
step, the nearest codebook entry is selected and subtracted from the residual, which
is then passed to the next codebook. The final discrete representation is obtained
by summing the selected embeddings from all codebooks and is fed into the decoder
for waveform reconstruction. Codebook entries are updated using exponential moving average updates with a decay
factor of 0.99. To prevent codebook collapse, unused entries are periodically
replaced by randomly sampled encoder vectors from the current batch. During
backpropagation, gradients are propagated through the quantization operation using
the straight-through estimator \cite{bengio2013estimating}.

\subsection{Discriminators}

To encourage high-fidelity and perceptually realistic reconstructions, adversarial
training is employed using multiple discriminators, following common practice in
neural codec training
\cite{encodec_defossez2022high,zeghidour2021soundstream,yang2023hifi}. Three discriminators are used: a Multi-Scale STFT (MS-STFT) discriminator, a
Multi-Scale Discriminator (MSD), and a Multi-Period Discriminator (MPD). The
MS-STFT discriminator operates on complex-valued short-time Fourier transforms at
multiple resolutions, where real and imaginary components are concatenated and
processed by a sequence of two-dimensional convolutional layers with increasing
temporal dilation. The MSD processes raw waveforms at multiple temporal scales,
while the MPD captures periodic structure by reshaping waveforms into
two-dimensional representations with different periods following by in~\cite{kong2020hifi}. All discriminators are configured with matched channel dimensions to balance their
contributions during adversarial training.
% \begin{figure*}[t]
%     \centering
%     \includegraphics[width=\linewidth]{latex/imgs/QC_final.png}
%     \caption{\textbf{Qualitative comparison of reconstructed spectrograms across different codecs.}
%     The figure visualizes mel-spectrograms of the same speech segment reconstructed by
%     (a) the natural reference, (b) our method, (c) DAC, and (d) EnCodec.
%     Low-frequency regions associated with prosodic and emotional cues are highlighted for comparison,
%     illustrating differences in temporal continuity and spectral stability across models.}
%     \label{fig:qualitative_comparison}
% \end{figure*}

\begin{figure*}[t]
    \centering
    \includegraphics[width=\linewidth]{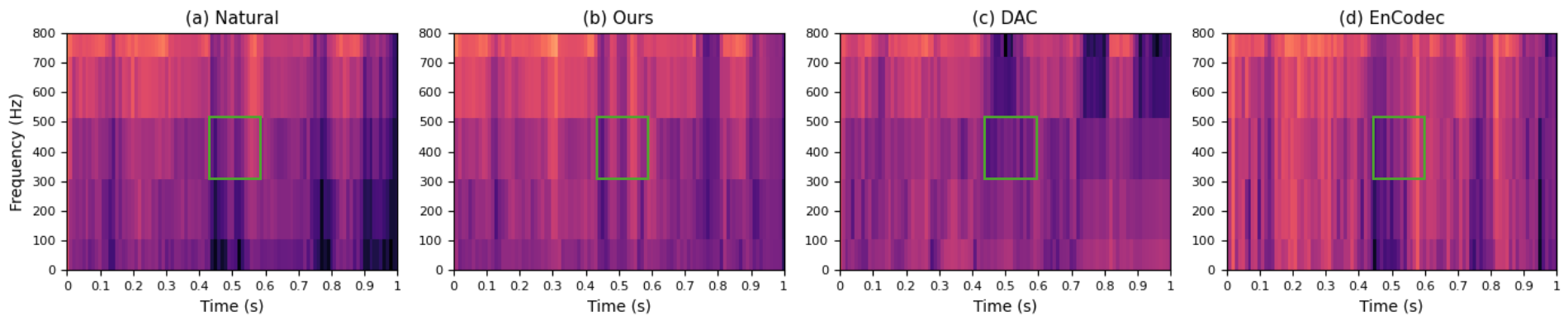}
    \caption{\textbf{Qualitative comparison of reconstructed spectrograms across different codecs.}
    The figure visualizes mel-spectrograms of the same speech segment reconstructed by
    (a) the natural reference, (b) our method, (c) DAC, and (d) EnCodec.
    Low-frequency regions associated with prosodic and emotional cues are highlighted for comparison,
    illustrating differences in temporal continuity and spectral stability across models.}
    \label{fig:qualitative_comparison}
\end{figure*}

\subsection{Training Objective}
\label{apx:training_objective}

The training objective combines reconstruction-oriented losses with adversarial supervision to ensure faithful waveform recovery, perceptual naturalness, and stable discrete quantization. Let $\mathbf{x}$ denote the input speech waveform and $\hat{\mathbf{x}}$ its reconstruction produced by the codec.

\paragraph{Quantization Commitment Loss.}
To stabilize residual vector quantization and prevent codebook collapse, we apply a commitment loss that penalizes the discrepancy between the encoder outputs and their quantized counterparts. Let $\mathbf{z}_j$ denote the residual vector at the $j$-th quantization stage and $\mathbf{z}_{q_j}$ the selected codebook embedding. The commitment loss is defined as
\begin{equation}
\mathcal{L}_{q}
=
\sum_{j=1}^{K}
\left\|
\mathbf{z}_j - \mathbf{z}_{q_j}
\right\|_2^2,
\tag{10}
\end{equation}
where gradients are applied only to the encoder outputs using a straight-through estimator.

\paragraph{Spectral Reconstruction Loss.}
To encourage spectral fidelity across multiple temporal resolutions, we compute reconstruction loss in the time--frequency domain using mel-spectrograms. Specifically, mel representations $M_i(\cdot)$ are extracted with STFT window sizes $2^i$ and hop sizes $2^i/4$, where $i \in \{5,\dots,11\}$. The spectral loss is defined as
\begin{equation}
\begin{aligned}
\!\!\!\!\!\!\!\!\!\!\!\!\!\mathcal{L}_{\mathrm{mel}}
&=
\sum_{i}
\Big(
    \left\| M_i(\mathbf{x}) - M_i(\hat{\mathbf{x}}) \right\|_1 \\
&\!+
    \left\| M_i(\mathbf{x}) - M_i(\hat{\mathbf{x}}) \right\|_2
\Big)
\end{aligned}
\tag{11}
\end{equation}
% \begin{equation}
% \mathcal{L}_{\mathrm{mel}} 
% =
% \sum_{i}
% \left(
% \left\| M_i(\mathbf{x}) - M_i(\hat{\mathbf{x}}) \right\|_1
% +
% \left\| M_i(\mathbf{x}) - M_i(\hat{\mathbf{x}}) \right\|_2
% \right).
% \tag{11}
% \end{equation}

\paragraph{Adversarial Loss.}
To enhance perceptual realism of reconstructed speech, adversarial supervision is introduced following prior neural codec frameworks~\cite{encodec_defossez2022high,yang2023hifi}. 
A set of discriminators $\{D_k\}_{k=1}^{K}$ operating at multiple temporal and spectral resolutions is employed, including multi-scale STFT, multi-period, and multi-scale waveform discriminators. The generator is optimized to produce reconstructions that are indistinguishable from real speech. 
Its adversarial loss is defined using the hinge formulation as
\begin{equation}
\mathcal{L}_{\mathrm{adv}}^{G}
=
\frac{1}{K}
\sum_{k=1}^{K}
\max\!\left(0,\, 1 - D_k(\hat{\mathbf{x}})\right),
\tag{12}
\end{equation}
where $\hat{\mathbf{x}}$ denotes the reconstructed waveform. The discriminators are trained to distinguish real speech $\mathbf{x}$ from generated samples $\hat{\mathbf{x}}$.
Their objective is given by
\begin{equation}
\begin{aligned}
\mathcal{L}_{\mathrm{adv}}^{D}
= \;&
\frac{1}{K}
\sum_{k=1}^{K}
\Big[
\max\!\left(0,\, 1 - D_k(\mathbf{x})\right) \\
&\quad +
\max\!\left(0,\, 1 + D_k(\hat{\mathbf{x}})\right)
\Big].
\end{aligned}
\tag{13}
\end{equation}
This adversarial formulation encourages high-fidelity waveform reconstruction while stabilizing training across multiple discriminative views.

\paragraph{Feature Matching Loss.}
To stabilize adversarial training and align intermediate representations, we additionally apply a feature matching loss over discriminator activations. Let $D_k^{(l)}(\cdot)$ denote the output of the $l$-th layer of discriminator $k$, with $L$ total layers. The feature matching loss is given by
\begin{equation}
\mathcal{L}_{\mathrm{feat}}
=
\frac{1}{K L}
\sum_{k=1}^{K}
\sum_{l=1}^{L}
\frac{
\left\|
D_k^{(l)}(\mathbf{x}) - D_k^{(l)}(\hat{\mathbf{x}})
\right\|_1
}{
\mathbb{E}\!\left[
\left\|
D_k^{(l)}(\mathbf{x})
\right\|_1
\right]
},
\tag{14}
\end{equation}
which encourages the generator to match real speech statistics across multiple abstraction levels.

\section{Qualitative Analysis}
\label{apx:qualitative}
Figure~\ref{fig:qualitative_comparison} provides a qualitative comparison of low-frequency mel spectrograms for natural speech, the proposed method, and two representative neural codecs. The analysis focuses on frequencies below 800 Hz, which are closely linked to prosodic and emotional cues~\cite{low_singh2022analysis,low_sharan2024emotion}, including fundamental frequency components, lower-order formants, and their temporal dynamics. These regions are particularly sensitive to quantization artifacts and thus offer an informative view of emotional structure preservation. A representative time–frequency region shared across methods (approximately 0.4–0.6 s and 300–500 Hz) is highlighted, where formant transitions and energy modulation are prominent. Natural speech exhibits smooth temporal continuity and coherent spectral evolution in this region. The proposed method closely preserves this behavior, maintaining stable and continuous spectral trajectories across frames. In contrast, DAC and EnCodec display more fragmented patterns, with abrupt temporal variations and reduced local coherence, indicative of quantization-induced disruption in low-frequency spectral structure. Overall, low-frequency spectral regions associated with prosodic and emotional cues show improved temporal continuity under the proposed method, suggesting reduced distortion of affectively salient structure during quantization. These qualitative observations are consistent with the quantitative gains in emotion consistency and perceptual naturalness, providing complementary evidence for the effectiveness of emotion-aware modeling.
\end{document}